\documentclass[12pt]{article}
\pdfoutput=1
\usepackage[a4paper]{geometry}
\usepackage[labelfont=bf]{caption}
\usepackage{subcaption}

\pdfoutput=1

\usepackage{comment}

\usepackage{jheppub,hyperref,float,array,adjustbox,mathtools,physics, xcolor}
\usepackage{graphicx,latexsym} 
\usepackage{amssymb,amsmath}
\usepackage{amsthm,lscape}
\usepackage{longtable,booktabs,setspace} 
\usepackage{url}
\usepackage{xcolor}

\definecolor{bittersweet}{rgb}{1.0, 0.44, 0.37}
\usepackage[bbgreekl]{mathbbol}

\usepackage{amsmath,amssymb,amsfonts,amsxtra, mathrsfs, makeidx,graphicx,amsthm,epsfig, bm,longtable,float, 
color,tikz,mathtools,xfrac,footnote,rotating, lscape, makecell, environ,mathtools, empheq, physics,cleveref,tensor,
slashed,subfiles,natbib,youngtab,multirow}

\usepackage{bbold}
\DeclareSymbolFontAlphabet{\mathbb}{AMSb}
\DeclareSymbolFontAlphabet{\mathbbl}{bbold}

\NewDocumentCommand{\fluxx}{ O{} O{} m m}{[\rho_{#3}^{#1}\rho_{#4}^{#2}]^{y_S}_{y_N}}
\newcommand{\SO}{\mathrm{SO}}
\newcommand{\USp}{\mathrm{USp}}
\newcommand{\Spin}{\mathrm{Spin}}
\newcommand{\Pin}{\mathrm{Pin}}
\newcommand{\SU}{\mathrm{SU}}
\newcommand{\U}{\mathrm{U}}
  
\newcommand{\textoverline}[1]{$\overline{\mbox{#1}}$}

\title{
\begin{center}
A double scaling for the 4d/3d reduction of $\mathcal{N}=1$ dualities
\end{center}
}

\author[a]{Antonio Amariti,}	
\author[a,b]{Andrea Zanetti}

\affiliation[a]{INFN, Sezione di Milano, Via Celoria 16, I-20133 Milano, Italy}
\affiliation[b]{Dipartimento di Fisica, Università degli studi di Milano, Via Celoria 16, I-20133}

\emailAdd{antonio.amariti@mi.infn.it}
\emailAdd{andrea.Zanetti@mi.infn.it}

\abstract{
In this paper we revisit the $S^1$ reduction of 4d $\mathcal{N}=1$ gauge theories, considering a double scaling on the 
radius of the circle and on the real masses arising from the global symmetries in the compactification. We discuss the implication of this double scaling for
SQCD with gauge algebra of ABCD type. We then show how our prescription translates 
in the reduction of the 4d superconformal index to the 3d squashed three sphere partition function. 
This allows us to derive the expected integral identities for the 3d dualities directly from the four dimensional ones.
This is relevant for the study of orthogonal SQCD, where the derivation from the 4d index is not possible in absence of the double scaling, because of a divergence due to a
flat direction in the Coulomb branch of the effective theory on the circle.
Furthermore, we obtain, for the even orthogonal case, a 3d duality with a quadratic fundamental monopole superpotential already discussed in the literature, that 
receives in this way an explanation from 4d.}

\begin{document}
\maketitle
\flushbottom
\allowdisplaybreaks 

%
%
%
%
%
%
\section{Introduction}
\label{sec:Introduction}
%
%
%
%
%
%
4d $\mathcal{N}=1$ gauge theories on $\mathbb{R}^{1,3}$ give rise in general to 3d $\mathcal{N}=2$ gauge theories when one dimension is compactified on a circle.
Such compactification has to be taken with some care if one has the goal to preserve an IR 4d duality, such as Seiberg duality \cite{Seiberg:1994pq} or its generalizations.
An efficient and general prescription has been described in full generality in \cite{Aharony:2013dha} (see also \cite{Dolan:2011rp,Imamura:2011uw,Gadde:2011ia,Niarchos:2012ah}  for an earlier result in this direction) for models with unitary and symplectic gauge groups, and in \cite{Aharony:2013kma} for models with orthogonal gauge groups. In the following we will refer to such prescription as the ARSW prescription. The ARSW prescription is based on the observation that the 4d duality is preserved if the finite size effects of the circle are taken into account in the compactification. 
While only the massless KK spectrum  on the circle is considered, the finite size effects modify the Coulomb branch through a superpotential for the so called KK monopole.
In the case of SQCD it corresponds to add to the usual superpotential for the BPS monopoles (expressed in terms of the roots of the associated  Lie algebra) an extra contribution coming from the affine root \cite{Davies:1999uw,Davies:2000nw,Kim:2004xx}. This last contribution, referred in \cite{Aharony:2013dha} as the KK monopole superpotential, is crucial to preserve the 4d duality. Indeed, it prevents the generation of new symmetries either anomalous in 4d  (e.g. axial symmetries) or topological in 3d, that would potentially destroy the duality.

From the 3d perspective such dualities can be thought as new effective dualities with a monopole superpotential turned on. The pure 3d limit is in general recovered by triggering a real mass flow\footnote{Such real mass flow can be accompanied by a Higgs flow that triggers a gauge symmetry breaking as well.}. Performing this flow leads to the ordinary 3d dualities originally studied in \cite{Aharony:1997gp,Karch:1997ux} and then generalized in various directions.
Observe that in some case one can engineer real mass flows  that maintain a monopole superpotential and in this last case new 3d effective dualities with 
monopole superpotentials have been constructed \cite{Benini:2017dud}. In most of the cases these superpotential are linear, but cases with quadratic deformations have been proposed as well (see also \cite{Amariti:2018gdc,Amariti:2019rhc}).

The ARSW prescription described so far has been translated into the language of localization, by studying the reduction of the 4d superconformal index \cite{Kinney:2005ej,Romelsberger:2005eg} to the 3d partition function on the (squashed) three sphere
\cite{Kapustin:2009kz,Jafferis:2010un,Hama:2010av,Hama:2011ea}. The problem in the mathematical literature corresponds to the hyperbolic limit of the identities among hypergeometric elliptic integrals \cite{rains2009}.
In this formulation of the problem the one-loop determinants that are obtained for the chiral and for the vector multiplets can be expressed in terms of  elliptic gamma functions. The KK reduction corresponds to an identity between a single elliptic Gamma function and an infinite amount of hyperbolic Gamma functions, \textit{i.e.} the one-loop determinants of the 3d multiplets.
Such infinite products reconstruct  the ``KK tower", where the 4d fugacities become the 3d real masses. Considering only the massless mode one can construct the relation between a single 4d one-loop determinant and a single 3d one-loop determinant. In the limit of small radius there is also a divergent pre-factor, that corresponds to the 4d gravitational anomaly.

By applying such limit to the integral identity that realizes a 4d duality on $S^3 \times \mathbb{R}$, an identity between 3d partition functions is obtained. The constraints imposed on the 4d fugacities, due to anomaly cancellation, reflect into constraints imposed by the KK monopole superpotential in the effective 3d picture.
If the divergent pre-factors cancel in the identity (as they should), one is left with a 3d integral identity. The cases where this integral identity is well-defined are interpreted as the realization of the effective duality from the localization perspective. Nevertheless, in general, the integral identity obtained from the procedure summarized above  is not guaranteed to hold true, as it is the case of orthogonal SQCD, because of the presence of a flat direction in the Coulomb branch, unlifted by the KK superpotential.

This problem has been noticed in \cite{Aharony:2013kma}, where a possible solution was suggested in terms of a double scaling limit on the radius and on the real masses. Such possibility 
would require to consider a vacuum state with some gauge holonomies and flavor fugacities on $S^1$ far away from the origin. This vacuum forces a gauge and a flavor symmetry  breaking and the 4d duality would be preserved as an effective duality if the magnetic gauge theory is considered in the corresponding dual vacuum (when it exists).
The superpotential involving the monopoles in such case should be affected, and the gauge symmetry breaking should imply the existence of a superpotential among the fundamental monopoles of the broken gauge sectors, hopefully lifting the flat direction in the Coulomb branch.
When such a scenario is realized, the procedure applied to the reduction of the SCI should give a finite and trustable result, that should allow to recover the pure 3d duality.

In this paper we give an explicit realization of a double scaling realizing the scenario discussed above. We focus on the case where a 4d $\mathcal{N}=1$ theory with a gauge group $G$ and a flavor symmetry $F$, with algebra of ABCD-type, is placed on $\mathbb{R}^{1,2} \times  S^1$ and both the gauge and the flavor symmetries are broken, respectively, into $G_1\times G_2 \subset G$ and $F_1 \times F_2 \subset F$. Such symmetry breaking pattern is realized by the compact scalar $\sigma$, corresponding to the fourth component of the field $A_\mu$ in four dimensions and by the scalars  $\sigma_F$ associated to the weakly coupled flavor symmetry (\textit{i.e.} the real masses).
By parametrizing $\sigma$ in the Cartan of the gauge symmetry $\U(1)^{\mathrm{rk}G}$, we consider a case where the vacuum corresponds to take
$\sigma_{1,\dots,\mathrm{rk}G_1}=0$ and  $\sigma_{\mathrm{rk}G_1 + 1,\dots,\mathrm{rk} G }=\frac{1}{2r_1}$ (where $1/r_1$ is the periodicity of the compact scalar $\sigma$ on the circle). 
Due to the symmetry of the point $1/2r_1$ on the circle we will often refer to this special locus as the mirror point, borrowing the nomenclature of \cite{Amariti:2015mva}.

Considering $\ell$ real scalars at a generic point on the circle $\sigma_1=\dots=\sigma_\ell \neq 0,\frac{1}{2r_1}$ the gauge symmetry is generically broken to $\U(\ell)$; on the other hand the mirror point is special, indeed, by placing the holonomies there,  the gauge symmetries of BCD-type algebra give rise to DCD-type algebras respectively.
We further consider a partial flavor symmetry breaking pattern, triggered by the real masses associated to the fundamental fields. 
This step is the crucial novelty of our approach. Indeed, in absence of such a flavor symmetry breaking pattern, the choice of the gauge configuration with the real scalars 
$\sigma_i$ at the origin is often the only possible one \cite{DiPietro:2016ond,Hwang:2018riu}. Breaking the flavor symmetry allows other stable configurations
where a gauge symmetry breaking pattern can be consistently realized on the circle.
We interpret the model obtained in this reduction as a 3d effective theory with  superpotential interactions  turned on between the monopoles of the $G_1$ and $G_2$ sectors. The 4d duality on $\mathbb{R}^{1,3}$  is then translated into a duality on $\mathbb{R}^{1,2}\times S^1$ if the dual  gauge symmetry breaking pattern is consistently chosen, giving rise to a model with two sectors with gauge group $\widetilde G_1$ and $\widetilde G_2$ respectively, interacting through their fundamental monopoles. 
Furthermore, we observe that these ``new'' dualities can be explained in terms of the ``pure'' 3d Aharony(-like) dualities: indeed one can locally dualize
\footnote{See \cite{Aharony:2013dha,Amariti:2015yea} for similar applications of local dualities  in this context.} $\widetilde G_1$ and $\widetilde G_2$  and obtain the dual electric model with  $G_1$ and $G_2$. This last observation is crucial to recover the canonical duality involving only the gauge group $G_1$ in the  double scaling limit. This consists of first dualizing only $\widetilde G_2$ and then to decouple the two sectors (in both the electric and  the dualized magnetic description) at the mirror point, both now with gauge group $G_2$, by sending the radius of the circle to zero.
When eliminating these sectors the only leftover are some singlets that originates from the 3d duality on $\widetilde G_2$. These singlets correspond to the monopole operators 
of the model with gauge group $G_1$, and they set to zero, in the chiral ring, the monopole operators of $\widetilde G_1$ through  a  superpotential interaction. 
In this way one arrives to the expected pure 3d limit and the final 3d duality is an Aharony(-like) duality as expected. 
While this procedure does not lead in principle to any new result from the field theory side for the ordinary\footnote{Actually it explains the 4d origin of 3d dualities with more exotic monopole superpotential, as we will see in the following.} 3d dualities, it becomes appealing when it is translated into the language of localization.
Indeed, the procedure discussed above lifts some of the flat directions in the Coulomb branch in the case of orthogonal SQCD, and it allows to derive the integral identity 
for 3d Aharony duality with orthogonal gauge group directly, without the aim of the standard trick discussed in 3d in \cite{Benini:2011mf} (and in 4d in \cite{Dolan:2008qi,Spiridonov:2011hf}).  

As a bonus we consider the ``extremal'' case of our prescription, \textit{i.e.} the case where the gauge symmetry is fully reconstructed at the mirror point, $\sigma_{1,\dots,\mathrm{rk}G}=\frac{1}{2r_1}$. This case does not give new insights for the ACD algebras, because the affine root, corresponding to the KK monopole superpotential, is exchanged with one ordinary roots, \textit{i.e.} one of the BPS monopoles. On the other hand  such exchange does not take place in the case with algebra of type $B$. Indeed, in this case the algebra at the special point is of type $D$ and the effective KK monopole  corresponds to $Y_{\Spin}$ or $Y^2$ depending on the global properties of the gauge group. In this way we arrive to a duality proposed  in \cite{Amariti:2018gdc} that receives then  a 4d explanation.
 
%
%
%
%
%
%
\section{4d/3d reduction: the ARSW prescription}
\label{sec:General}
%
%
%
%
%
%
It this review section we discuss the main aspects of the ARSW prescription designed to derive 3d dualities starting from 4d parents.
The key observation is that when considering the 4d dynamics with one compact circular direction there are new effects that arise at finite size and that vanish in the strict zero radius limit. Such effects modify the compact Coulomb branch on the circle, through a superpotential, referred to as KK monopole superpotential. If the zero modes in the KK monopole background arise only from the vector multiplet, the KK monopole superpotential is generated, and it constrains the symmetries of the effective 3d theory, namely by breaking the ones that are anomalous in 4d and in general the new global symmetries  that are generically generated in 3d.
A 4d duality is then preserved at finite size if such effects are included. The pure 3d limit is obtained by  a real mass flow, that removes the effect of the KK monopole, giving rise to an ordinary 3d duality.
Actually this last step is not necessary in the prescription if the KK monopole superpotential does not completely lift the Coulomb branch. This is for example the case of  orthogonal SQCD with vectors. In this last case the pure 3d limit is obtained by mapping the unlifted 
Coulomb branch coordinates between the dual phases on the circle. Focusing on one region close to the origin of the Coulomb branch in one phase requires focusing on a gauge symmetry breaking region in the dual phase. Removing the KK monopole superpotential in this case can be done without further real mass flows, and one ends up with the expected pure 3d duality at zero size.

More concretely let us consider a 4d $\mathcal{N}=1$ pure gauge theory on a circle with a vector multiplet $V$. The fourth component of the gauge field $A_4$ becomes a compact real scalar $\sigma$. In this way  one reconstructs a 3d $\mathcal{N}=2$ vector multiplet with a compact Coulomb branch.
At a generic point of the Coulomb branch the non-abelian gauge group $G$  is broken to $\U(1)^{\mathrm{rk}G}$ and the  scalars $\sigma_{i=1,\dots,\mathrm{rk}G}$
refer to the Cartan subgroup of $G$. 
The Coulomb branch is usually parametrized by the coordinates 
\begin{equation}
Y_i = e^{\Sigma_i} = e^{\frac{4 \pi \sigma_i}{g_3^2}+i\varphi_i},
\end{equation}
where the compact scalar $\varphi_i$ represent the dual photons at a generic point of the Coulomb branch.
The 3d gauge coupling $g_3$ is obtained from the 4d one through the relation $g_4^2 = r_1 g_3^2$
where $g_4$ is obtained from the 4d holomorphic scale 
$\Lambda_{holo}^{b} \simeq e^{-\frac{1}{g_4^2}}$.

Microscopically (some of) the Coulomb branch variables are related to monopole operators \cite{Borokhov:2002ib,Borokhov:2002cg,Kapustin:2005py}.
The insertion of a monopole operator at a point removes a ball around the point and puts one unit
of flux through its surface.
This has the same effect as the insertion of $\Sigma$ in the functional integral\footnote{The insertion of a monopole at a point $x_0$ generates a magnetic flux on the sphere surrounding $x_0$
and by considering $\sigma$ pointing in the direction of the flux one has 
$\lim_{x \rightarrow x_0} \sigma(x) \rightarrow \infty$.}.

These monopoles act as instantons in 3d generating an effective superpotential on the Coulomb branch. For a pure gauge theory on the circle the superpotential has the form
\cite{Davies:1999uw,Davies:2000nw,Kim:2004xx}
\begin{equation}
\label{Wmon}
W_{mon} = W_{BPS} + W_{KK},
\end{equation}
where 
\begin{equation}
\label{WBPS}
W_{BPS} = \sum_{j=1}^{\mathrm{rk}G} \frac{2}{\alpha_j^2} e^{\alpha_j^* \cdot\Sigma}
\end{equation}
and
\begin{equation}
\label{WKK}
W_{KK} =  \frac{2 \eta }{\alpha_0^2} e^{ \alpha_0^* \cdot\Sigma}
\end{equation}
In these formulas $\alpha_i$ are the simple roots of $G$ and $\alpha_i^*$ are the co-roots of $G$. Furthermore,
$\alpha_0$ is the affine root of $G$  and $\alpha_0^*$ is the affine co-root of $G$.
The contribution (\ref{WBPS}) survives in the 3d limit, where the circle has zero size, while the superpotential 
(\ref{WKK}) vanishes as $r_1 \rightarrow 0$ because  $\eta \equiv e^{4 \pi/(r_1 g_3^2)}$.
Physically the superpotential (\ref{Wmon}) reproduces the vacua of the 4d parent theory while in the 3d limit the vacua are sent to infinity.

At this point of the discussion one needs to include the effect of matter fields. Depending on their representation
some of the matter fields induce new fermionic zero modes that can prevent the generation of the monopole superpotential.
Focusing on cases where $W_{BPS}=0$  while $W_{KK} \neq 0$ one can construct new effective dualities on $\mathbb{R}^{1,2} \times S^1$ starting from 
4d dual pairs.
Such dualities have the same structure of the 4d parents except the fact that they are equipped by  a $W_{KK}$ monopole superpotential 
in the electric and in the magnetic phase. The presence of such a superpotential prevents the generations in 3d of symmetries that would spoil the 
4d duality on $S^1$. The pure 3d limit consists of removing the contribution of the $W_{KK}$ superpotential consistently in the electric and in the magnetic phase.

This step requires some care, because depending on the dual pair different prescriptions are necessary. For example in some case it may be necessary to 
study a real mass flow, while in other cases a further Higgs flow (triggered by the real scalar in the vector multiplet) must be considered. On the other hand if there are flat directions, in the Coulomb branches of the effective dual pairs, at finite size it is
still possible to obtain the 3d limit without any real mass or Higgs flow but by focusing on the correct dual regions in the moduli space and then sending the radius of $S^1$ to zero.
This last possibility has been explicitly studied in \cite{Aharony:2013kma} for orthogonal SQCD with vectors. While the final result provides the expected 3d dualities the prescription cannot be translated immediately in the language of localization, \textit{i.e.} in reduction of the 4d identity between the SCI of the dual sides of the duality to the expected identity between the 
3d partition functions.
This is the main motivation behind our analysis: we aim to modify the prescription of ARSW in order to follow the various steps of the reduction from the perspective of localization as well.

Before doing that let us review in more details the case of $so(N)$ SQCD with $N_f$ vectors.
In this case a $\mathbb{Z}_2$ subgroup of the center is left unbroken by the matter fields and the global properties are relevant. In 4d there are three possible dualities, one between $\SO(N)_+$ gauge groups and the other two involving a $\Spin(N)$ gauge group on one side and a $\SO(N)_-$ gauge group on the other.

In the $\Spin(N)$ case all the monopoles carry charges corresponding to the co-roots of the gauge group.
The minimal monopole is  $Y_{\Spin}$ related to the Coulomb branch variables by
\begin{equation}
Y_{\Spin} \propto e^{2\left(\frac{4 \pi \sigma_1}{g_3^2} +  i \varphi\right)}.
\end{equation}
The KK monopole is 
\begin{equation}
Z = e^{\frac{4 \pi}{g_3^2}(\sigma_1+\sigma_2) + i (\varphi_1+\varphi_2)}.
\end{equation}
When the group is $\SO(N)_{\pm}$ there are also monopoles that do not 
correspond to co-roots. The minimal monopole is $Y$, defined such that 
$Y^2 = Y_{\Spin}$.
The monopole superpotential is written in terms of the Coulomb branch coordinates
\begin{equation}
Y_i= e^{\frac{4 \pi (\sigma_i-\sigma_{i+1})}{g_3^2}+i (\varphi_i-\varphi_{i+1})},
\end{equation} 
holding for both the $B_N$ and  $D_N$ case with  $i=1,\dots,N-1$ and 
\begin{equation}
Y_{N}= e^{\frac{4 \pi (\sigma_{N-1}+\sigma_{N})}{g_3^2}+i (\varphi_{N-1}+\varphi_{N})}
\end{equation} 
or
\begin{equation}
Y_{N}= e^{\frac{8 \pi \sigma_{N}}{g_3^2}+i 2\varphi_{N}}
\end{equation} 
for $D_N$ and $B_N$ respectively.
In presence of massless vectors the fermionic zero-mode counting lifts the BPS monopole superpotential and one is left with the 
contribution of the KK monopole. However, a one dimensional Coulomb branch is leftover, parametrized by $Y_{\Spin}$ or $Y$.
This flat direction is non trivially mapped across the dual phases and by carefully considering this relation the pure 3d limit can be recovered.
In general one can reduce the $\SO(N)_+$ case, finding a $\SO(N)_+$ duality on $S^1$ and then reduce to a pure 3d $\SO(N)$ duality. 
From this duality one can 
construct various 3d dualities with different global structures (involving $\mathrm{O}(N)_\pm$, $\Spin(N)$ or $\Pin(N)$\footnote{More precisely $\Pin(N)_{\pm}$ see \cite{Cordova:2017vab}.}) by discrete gaugings of the $\mathbb{Z}_2$ charge 
conjugation symmetry $\mathbb{Z}_2^C$ and the magnetic $\mathbb{Z}_2^M$ symmetry that comes from the non-trivial center of the gauge group and that charges the sign of the fundamental monopole.

%
%
%
%
%
%
\section{A different prescription: double scaling limit}
\label{sec:double}
%
%
%
%
%
%

Despite the fact that the ordinary $\mathrm{O}(N)_+$  duality   \cite{Benini:2011mf,Aharony:2011ci} is recovered by the ARSW prescription, here we are interested to modify the prescription in order to lift the flat direction in the Coulomb branch on $S^1$. 
Such a lift is necessary from the viewpoint of localization in order to mimic the prescription when reducing the 4d index to the 3d partition function. As suggested in \cite{Aharony:2013kma}
it should be possible to remove the flat direction by considering a double scaling limit, when the small radius limit is taken  while scaling some scalars in the background 
vector multiplets (\textit{i.e.} 3d real masses) to be large.

This possibility may require also to scale some of the gauge holonomies, \textit{i.e.} a partial Higgsing of the gauge group on $S^1$. In other words, we look for a 
vacuum for the gauge holonomies consistent with the background that has been modified by the non-trivial choice of the flavor holonomies. For simplicity, we focus on the case of two packages of holonomies, one at the origin of the circle and one at the mirror point.
In this way one arrives to a product group gauge theory $G_1 \times G_2$ with a split flavor symmetry $F_1 \times F_2$. Preserving the duality on the circle requires to correctly identify the dual split of the gauge symmetry $\tilde G \rightarrow \tilde G_1 \times \tilde G_2$.
The pure $3d$ limit then requires to decouple the sectors at $1/2r_1$ preserving the duality. In the examples considered below we observe that this last step can be done in each case by a local 3d duality on the sector identified by $\tilde G_2$. This indeed allows to decouple in the $r_1 \rightarrow 0$ limit the  sectors at $1/2r_1$ and one is left with the expected pure 3d duality.
While the procedure is motivated by the analysis of orthogonal gauge theories we can observe that it can be used for more general gauge theories and indeed  we will study in the explicit examples below also the cases of ordinary $\U(N)$ and $\USp(2N)$ SQCD dualities.

We conclude this section by studying the split of $G$ into $G_1$ and $G_2$ on the Coulomb branch.
Let us split $\SO(2N + 1)$ by choosing $N - \ell$ gauge holonomies at zero and
$\ell$ at $\frac{1}{2r_1}$.
Defining 
\begin{align}
& \chi_i \equiv \sigma_{\ell+i}  & i=1,\dots,N - \ell \nonumber\\
& \rho_i \equiv - \sigma_{\ell-i+1} +\frac{1}{2r_1} & i=1,\dots,\ell \nonumber\\
\end{align}
the coordinates $Y_i$ become
\begin{eqnarray}
&& Y_i \rightarrow \tilde{Y}^{(D)}_{\ell-i} \quad i=1.\dots,\ell-1 \nonumber\\
&& Z \rightarrow \eta^{-1}/{\tilde Y^{(D)}_{\ell}} \nonumber\\
&& Y_{\ell} \rightarrow \left(\frac{1}{\eta \tilde Y^{(D)} \tilde Y^{(B)}}\right)^{\frac{1}{2}} \nonumber\\
&& Y_{\ell+i}\rightarrow \tilde{Y}^{(B)}_i \quad i=1,\dots,N - \ell.
\end{eqnarray}
The monopole superpotential becomes
\begin{eqnarray}
W_{2N+1} 
= 
\sum_{i=1}^{N - \ell - 1}\frac{1}{ \tilde{Y}^{(B)}_i}+ \frac{2}{ \tilde{Y}^{(B)}_{N - \ell}}
+
\sum_{i=1}^{\ell - 1}\frac{1}{ \tilde{Y}^{(D)}_i}+ \frac{1}{ \tilde{Y}^{(D)}_{\ell }}
+
\sqrt{\eta} \, \tilde Y^{(D)} \tilde Y^{(B)}.
\end{eqnarray}
In the $\SO(2N)$ case an analogous split holds such that the superpotential becomes 
\begin{eqnarray}
W_{2N}
= 
\sum_{i=1}^{N - \ell - 1}\frac{1}{ \tilde{Y}^{(D)}_i}+ \frac{1}{ \tilde{Y}^{(D)}_{N - \ell}}
+
\sum_{i=1}^{\ell - 1}\frac{1}{ \tilde{Y}^{(D)}_i}+ \frac{1}{ \tilde{Y}^{(D)}_{\ell}}
+
\sqrt{\eta} \,   \tilde Y^{(D)} \tilde Y^{(D)}.
\end{eqnarray}
In presence of $F$ vectors the leftover superpotential is just  
$\sqrt{\eta}  \tilde Y^{(D)} \tilde Y^{(B/D)}$  
 in the $\SO(2N+1)/\SO(2N)$ cases or 
  $\sqrt{ \eta \tilde Y_{\Spin}^{(D)} \tilde Y_{\Spin}^{(B/D)}}$
 in the $\Spin(2N+1)/\Spin(2N)$ cases. This lifts the Coulomb branch, and we are left with two orthogonal gauge theories, interacting through such monopole superpotential.
Observe that in this case we keep $F - h$ real masses at zero, while we scale $h$ real masses as $1/2r_1$. 
The model obtained so far on the circle can be sent to a pure 3d theory by sending the radius to zero.
In this limit $\eta \rightarrow 0$ and the final picture consists of two decoupled models (at infinite distance in the Coulomb branch). On one side we have a 
$\SO(2(N - \ell)+1)$ or $\SO(2(N -\ell))$ theory with $F - h$ vectors. On the other side we have a $\SO(2\ell)$ gauge theory with $h$ vectors.
 
While the procedure discussed so far is generic and can be applied in principle to any gauge theory, we will study in the various examples the prescription necessary to preserve the duality in this limit by focusing on the various cases. In general, we need to find the split of the dual gauge group, and then we must consistently decouple the sectors at infinite distance in the Coulomb branch such to preserve the duality.

%
%
%
\section{Double scaling and localization: from  $\mathcal{I}_{4d}$ to $Z_{S^3_b}$}
\label{subsec:SCItoZ}
%
%
%

The presence of dualities between theories implies identities between the corresponding partition functions. As previously discussed, in the context of the reduction of 4d dualities to 3d ones, it is necessary to first define the theories on $S^1 \times S_b^3$ and then to perform a dimensional reduction along the thermal circle to recover a 3d duality from a 4d one. From the localization perspective this procedure relates $ Z_{S^1 \times S_b^3} $ to $S_b^3$ partition functions and depending on the details of the prescription for the dimensional reduction, one ends with a gauge theory with possibly non-trivial matter content and interactions after decoupling the massive KK modes for the fields.
Then, from physical considerations it follows that

\begin{equation}
    \label{eq:KK_tower}
    Z_{S^1 \times S_b^3} \approx \prod_{KK modes} ^ {\infty} Z_{S_b^3}.
\end{equation}

In order to define a supersymmetric theory on $S^1 \times S_b^3$, with $r_1$ the radius of $S^1$ and $r_3$ the radius of $S_b^3$, we require the symmetry group to include at least a $\U(1)_R$ symmetry \cite{Festuccia:2011ws}. The isometries of the background define a $\SU(2)_1 \times \SU(2)_2 \times \U(1)$ symmetry group with generators $j_1, j_2$ and $\Delta$ respectively. The supersymmetric partition function for a supersymmetric field theory on $S^1 \times S_b^3$ is known to be proportional to the 4d sphere index of the theory, with the difference between the two given by the supersymmetric Casimir energy \cite{Assel:2015nca}. This term can be ignored when considering the dimensional reduction along $S^1$, being proportional to the size of the thermal circle. \\
Turning on fugacities for the charges of the theory we can define the 4d index as \cite{Kinney:2005ej, Romelsberger:2005eg}

\begin{equation}
    \label{eq:4d_ind}
    \mathcal{I}_{4d} = \mathrm{Tr} (-1) ^F e ^{- \beta \delta} p^{j_1 + j_2} q^{j_1 - j_2} (pq)^{R/2}\prod_k v_k ^ {e_k},
\end{equation}

\begin{equation}
    \delta \coloneqq \{\mathcal{Q}, \mathcal{Q}^\dagger\} = \Delta - 2j_1 - \frac{3}{2}R,
\end{equation}

which coincides with the superconformal index when the theory enjoys superconformal symmetry, with $\Delta$ associated to the conformal dimension. Equivalently, \cref{eq:4d_ind} can be interpreted as the partition function of the theory on $S^1 \times S_b^3$, squashed by the fugacities $p$ and $q$ parametrizing the angular momenta of the three-sphere, with squashing parameter $b^2 = \frac{\log(p)}{\log(q)}$, and twisted periodic boundary conditions for the fields due to the presence of additional fugacities \cite{Imamura:2011wg}.
The squashed three-sphere background requires $p$ and $q$ to be parametrized as\footnote{We consider $S_b^3$ with unit radius for simplicity, it is straightforward to introduce explicitly $r_3$ in all the expressions, see \cite{Aharony:2013dha} for instance. Notice that $r_1$ in our expressions is a dimensionless parameter, as it should be, corresponding to the dimensionless ratio $\frac{r_1}{r_3}$ for $r_3 \neq 1$.}
\begin{equation}
    p = e^{2 \pi i \tau},  \quad q = e^{2 \pi i \sigma},
\end{equation}
\begin{equation}
    \tau = \frac{\beta}{2 \pi}\omega_1, \quad \sigma = \frac{\beta}{2 \pi}\omega_2, \quad \beta = 2\pi r_1, \quad \frac{\tau + \sigma}{2} \equiv r_1 \omega.
\end{equation}
We parametrize the additional gauge and flavor fugacities respectively as 

\begin{equation}
    z_j = e^{2 \pi i u_j},  \quad v_k = e ^{2\pi i m_k}.
\end{equation}
It is customary to define a democratic basis for the fugacities of each field labeled by $a = 1,..., F$, which mixes R and flavor charges,
\begin{equation}
    (pq)^{R_a/2} \prod_{k} v_i^{e_k^a} \equiv y_a \equiv e^{2\pi i \Delta_a},
\end{equation}
together with a balancing condition 
\begin{equation}
    \sum_{a=1}^{F} \Delta_a = f(\tau,\sigma,F,N,...),
\end{equation}
constraining the charges so to encode the symmetries of the theory.
The sphere index for a 4d gauge theory with gauge group $G$ and $F$ flavors transforming in the representation $R_G$ under $G$ can be written as an elliptic hypergeometric integral of elliptic Gamma functions \cite{Dolan:2008qi}
\begin{equation}
    \label{eq:ind_gauge}
    \mathcal{I}_{4d}(y;p,q) = \frac{(p;p)_\infty^{\mathrm{rk}G} (q;q)_\infty^{\mathrm{rk}G}}{|W|} 
    \int \prod_{i=1}^{\mathrm{rk}G} \differential u_i \frac{\prod_{a = 1} ^ F \prod_{\rho_a \in R_G} \Gamma_e((pq) ^{R_a/2} z^{\rho_a(u)} v^{\rho_f(m)_a}; p, q)} 
    {\prod_{\alpha \in \Delta_+}  \Gamma_e( z ^{\alpha(u)} ; p, q)\Gamma_e( z ^{- \alpha(u)} ; p, q)},
\end{equation}
with $\rho$ and $\alpha$ represent the weights and the roots relative to the representation under which the matter and vector fields transform, $|W|$ is the cardinality of the Weyl group of $G$ and 
\begin{equation}
    (z;q)_\infty \coloneqq \prod_{k=0}^\infty \left(1 - z q^k \right),
\end{equation}
\begin{equation}
    \Gamma_e(z;p,q)\coloneqq \prod_{j,k = 0}^{\infty} \frac{1 - p^{j + 1} q^{k + 1}z^{-1}}{1 - p^j q^k z}.
\end{equation}
We also introduce the modified elliptic Gamma function
\begin{equation}
    \tilde \Gamma_e(u;\tau,\sigma) \coloneqq \prod_{j,k = 0}^{\infty} \frac{1 - e^{2\pi i((j + 1)\tau + (k + 1)\sigma - u)}}{1 - e^{2\pi i (j\tau + k\sigma + u)}},
\end{equation}
and suppress the explicit dependence on $\tau$ and $\sigma$, so that the elliptic Gamma function will be denoted simply by $\tilde \Gamma_e(u)$.
If the gauge group possesses some abelian factors, an extra FI term with parameter $\xi_4$ can be introduced in the index for each $\U(1)$ factor. When the theory is defined on $S^1 \times S_b^3$, quantization conditions must be imposed to preserve invariance under large gauge transformations along the compact circle. As the component of the gauge field along $S^1$ becomes periodic, namely $A_4 \sim A_4 + \frac{1}{r_1}$, the FI parameter must satisfy the quantization condition \cite{Aharony:2013dha}
\begin{equation}
    \label{eq:FI_par}
    \xi_4 = \frac{n}{4\pi^2}, \quad n \in \mathbb{Z}.
\end{equation}
This corresponds to an insertion of a $z^n$ factor in the 4d index in \cref{eq:ind_gauge}, with $z$ the holonomy parametrizing the abelian factor.

The physical statement expressed by \cref{eq:KK_tower} requires the elliptic Gamma functions to admit an infinite product expansion of (squashed-)three sphere partition functions for the KK tower of 3d chiral multiplets. 

The partition function of a 3d chiral multiplet on a squashed $S_b^3$ with squashing parameter $b$ and unit radius \cite{Hama:2011ea} is expressed in terms of Rains' hyperbolic Gamma functions \cite{rains2009}. Let us introduce $\omega_1  = i b$, $\omega_2 = i b ^{-1}$ and $\omega = \frac{\omega_1 + \omega_2}{2}$. Assuming $\Im\left(\frac{\omega_2}{\omega_1}\right) > 0$, the hyperbolic Gamma function can be defined in terms of an infinite product representation of the quantum dilogarithm $\psi_b(z)$\footnote{The function $\psi_b(z)$ is related to the Faddeev's quantum dilogarithm $e_b(z)$ in \cite{Faddeev:2000if} by $\psi_b(z)=e_b(-iz)$.} (see for instance Corollary 6 of \cite{NARUKAWA2004247})

\begin{equation}
\label{eq:hyp_gamma_id}
\exp{\frac{\pi i}{2 \omega_1\omega_2}\left(z^2 + \frac{\omega_1^2 + \omega_2 ^ 2}{12}\right)} \Gamma_h(iz + \omega;\,\omega_1,\,\omega_2) = \psi_b(z),
\end{equation}
with
\begin{equation}
    \psi_b(z) \coloneqq \frac{\left( - e^{\frac{2 \pi z }{\omega_2} + i \pi \frac{\omega_1}{\omega_2}}; e^{2\pi i \frac{\omega_1}{\omega_2}}\right)_\infty}{\left(- e^{\frac{2 \pi z }{\omega_1} - i \pi \frac{\omega_2}{\omega_1}}; e^{-2 \pi i \frac{\omega_2}{\omega_1}}\right)_\infty}.
\end{equation}
In the following we suppress conventionally the explicit dependence on $\omega_1$, $\omega_2$ in $\Gamma_h(iz + \omega;\, \omega_1,\, \omega_2)$.

The function $\psi_b(z)$ satisfies the inversion formula 

\begin{equation}
    \psi_b(z)\psi_b(-z) = \exp{\frac{\pi i}{\omega_1\omega_2}\left(z^2 + \frac{\omega_1^2 + \omega_2 ^ 2}{12}\right)},
\end{equation}
from which also the inversion formula 
\begin{equation}
    \Gamma_h(x + \omega)\Gamma_h(x - \omega) = 1
\end{equation}
for the hyperbolic Gamma functions immediately follows.

The 3d partition function on $S_{b}^3$ for a gauge theory with gauge group $G$ and $F$ flavors can then be written. It corresponds to a matrix integral over the scalar $\sigma$ of the 3d vector multiplet. Each matter field contributes to the partition function at the 1-loop level with an hyperbolic Gamma function, while the vector multiplet contributes both classically with a $e ^ {\frac{\pi i \sigma_i^2}{\omega_1 \omega_2}}$ term if a CS term is present and at the 1-loop level with $\Gamma_h(\alpha(\sigma))^{ - 1}$, after combining with the integration measure.

Moreover, when the gauge group possesses abelian factors, also a Fayet-Iliopoulos term is allowed in the partition function. This can be interpreted as a real mass parameter for the topological symmetries $\U(1)_J$ of the theory. The FI parameter $\xi_3$ can be related to $\xi_4$ \cite{Aharony:2013dha}:
\begin{equation}
\label{FI3d}
    \xi_3 = - 4 \pi^2 r_1 \xi_4 = n r_1 \, \omega_1 \omega_2, \quad n \in \mathbb{Z}.
\end{equation}
All in all, the 3d partition function is
\begin{equation}
    \label{eq:3dparf}
    \mathcal{Z}(\mu; \xi_3) = \frac{1}{|W|}\int \prod_{i=1}^{\mathrm{rk}G}\frac{\differential \sigma_i}{\sqrt{-\omega_1 \omega_2}} 
    e ^ {\frac{2 \pi i \xi_3 \sigma_i}{\omega_1 \omega_2}}   e ^ {\frac{\pi i \sigma_i^2}{\omega_1 \omega_2}} \frac{\prod_{a=1}^{F} \prod_{\rho \in R_G} \Gamma_h(\rho_a(\sigma) + \mu_a)}{\prod_{\alpha \in \Delta_+} \Gamma_h(\alpha(\sigma))\Gamma_h(-\alpha(\sigma))},
\end{equation}
where $\mu_a \coloneqq \rho_f(m)_a + \omega R_a$ are understood as real mass parameters incorporating the $R$ and flavor fugacities of the theory. 

Defining for convenience 
\begin{eqnarray}
    t_n (u)  \coloneqq && \frac{2 \pi}{i \beta} \left(u - \frac{\tau + \sigma}{2} + n \right) = -i \left(\frac{u + n}{r_1} - \omega\right) , \quad n \in \mathbb{Z}
\end{eqnarray}
we can relate elliptic and hyperbolic Gamma functions.
The elliptic Gamma functions satisfy the identity \cite{NARUKAWA2004247}

\begin{equation}
\label{eq:el_gamma_id}
   \tilde \Gamma_e(u) =  e^{i \pi Q(u)} \psi_b (t_0(u)) \prod_{n=1}^{\infty} \frac{\psi_b (t_n(u))}{\psi_b ( - t_{-n}(u))},
\end{equation}
where 
\begin{eqnarray}
    Q(u) \coloneqq -\frac{u^3}{3\tau \sigma} + \frac{\tau + \sigma - 1}{2\tau \sigma} u^2 - \frac{\tau^2 + \sigma^2 +3\tau\sigma - 3\tau - 3\sigma + 1}{6 \tau \sigma} u \\
     - \frac{1}{12}(\tau + \sigma - 1)(\tau^{-1}+\sigma^{-1} - 1) + \frac{\left(u - \frac{\tau + \sigma}{2}\right)^2}{\tau \sigma} - \frac{\tau^2 + \sigma^2}{12\tau \sigma}.
\end{eqnarray}
Upon substitution of \cref{eq:hyp_gamma_id} in \cref{eq:el_gamma_id} we get, defining $\mathrm{sign}(0) = 1$,
\begin{equation}
    \label{eq:el_gamma_hyp}
    \tilde \Gamma_e(u) =  e^{i \pi Q(u)} \prod_{n=-\infty}^{\infty} e^{- \mathrm{sign}(n) \frac{\pi i }{2\omega_1 \omega_2} \left( \left(\frac{u + n}{r_1} - \omega \right)^2 - \frac{\omega_1 ^2 + \omega_2 ^ 2}{12} \right)} \Gamma_h \left(\frac{u + n}{r_1} \right),
\end{equation}
which is the precise formulation of the physical KK reduction in \cref{eq:KK_tower}.

The exponential terms within the product act as regulators in the small $r_1$ limit, precisely cancelling out the asymptotic divergent behavior of the hyperbolic Gamma functions \cite{rains2009}:
\begin{equation}
        \Gamma_h(z;\omega_1,\omega_2)\underset{|z| \to \infty}{\sim} \exp{ \mathrm{sign}(z)\frac{\pi i }{2\omega_1 \omega_2} \left( \left(z - \omega \right)^2 - \frac{\omega_1 ^2 + \omega_2 ^ 2}{12} \right)}.
\end{equation}

The reduction of $S^1 \times S_b^3$ along the thermal circle lifts the massive KK modes on the $S^1$ leaving only zero modes in the effective theory. 

To obtain a non-trivial 3d theory in the reduction process we must assign an appropriate scaling in the fugacities of the 4d theory. The standard parametrization assigns a linear dependence on $r_1$ on all the fugacities. For this reason let us discuss first the $r_1 \to 0$ of the elliptic Gamma function with $u =  x r_1$. This is the so-called hyperbolic limit in the mathematical literature \cite{Van_Diejen_2005, rains2009}.

Taking the 3d limit $r_1 \to 0$ in \cref{eq:el_gamma_hyp} the massive KK modes decouple and only the $n=0$ term survives, giving

\begin{equation}
\tilde \Gamma_e(u) \underset{r_1 \to 0}{=}  e^{-\frac{i \pi (x - \omega) }{ 6 r_1 \omega_1\omega_2} + O(r_1)} \Gamma_h(m),
\end{equation}
where the Casimir energy $\mathcal{O}(r_1)$ enters linearly and thus does not contribute in the hyperbolic limit. In this limit a divergent contribution appears which can be understood as a mixed $\U(1)_R$ (and flavor)-gravitational anomaly term, as discussed in \cite{Aharony:2013dha}. It can be interpreted as a non-trivial background for the $\U(1)_R$ symmetry. This divergent term can be neglected as long as dualities are concerned, upon proving that the same contribution arises in both theories. This is in fact expected, due to 't Hooft anomaly matching arguments between dual theories. Such fact will reveal to be crucial below, where a matching between these terms in the dual phases will select the vacuum around which performing the 3d reduction, alongside with a non-trivial background for the flavor symmetries.

The standard hyperbolic limit procedure, in which a linear scaling is assigned to all the fugacities, has the net effect of reducing the 4d theory with zero real mass background around vanishing holonomies. This is the starting point for the ARSW prescription \cite{Aharony:2013dha} to engineer 3d dualities from 4d dualities directly from the partition functions. When applying this prescription to SQCD both the electric and the magnetic theories are first expanded around the gauge holonomies configuration $\sigma = (0,\dots,0)$. It is then possible to explore different vacua in the Coulomb branch by implementing a real mass flow for the flavor fugacities as discussed in \cite{Aharony:2013dha} and recover 3d dualities. When a mass flow is implemented, the vacua in the electric and the magnetic theory are not trivially related anymore. A vacuum at vanishing holonomies in the electric theory could be related to a non-trivial point in the Coulomb branch in the magnetic side. 

However, this procedure fails to work in some cases. For instance, it is not possible to obtain 3d dualities from the reduction of the partition function with this prescription, for theories where the Coulomb branch of the 3d theory is not completely lifted near the $\sigma = 0$ vacuum by monopole-induced superpotential terms. This is precisely the case for orthogonal SQCD. The consequences of such flat directions in the Coulomb branch on the three-sphere partition function have been discussed in \cite{ArabiArdehali:2015ybk}. In this situation taking the hyperbolic limit produces a divergent partition function in 3d, due to the index being insensitive to the unlifted direction on the Coulomb branch, and makes it impossible to recover a 3d duality from localization \cite{Aharony:2013kma}. More generally, it is not always the case that the point $\sigma = 0$ is a stable vacuum of the theory or other vacua can be possible \cite{ArabiArdehali:2015ybk,Hwang:2018riu}.

For this reason we modify the ARSW prescription by defining a ``double scaling limit'', which turns on a fixed non-trivial background for the real masses, while compactifying the thermal circle. When studying the theory on this background some or all of the KK matter fields zero modes can become massive and  get lifted. Depending on the choice of the background, we can thus reduce to 3d theories with different matter content, with respect to the original one. This has the effect of selecting a different vacuum when reducing the theory on $S_b^3$, in which correspondingly different monopoles superpotentials are generated on the Coulomb branch.

To this aim we modify the standard parametrization of the fugacities of the hyperbolic limit by fixing a background for the real masses and a generic non-zero vacuum for the holonomies. Equivalently, we perform a double scaling limit, in which the real masses are taken to be large with a $1/r_1$ scaling, while the radius of the $S^1$ goes to zero. Therefore, we parametrize the 4d fugacities as
\begin{equation}
    \label{eq:double_scal_pr}
    u_i = \sigma_i^* + \sigma_i {r_1}  \quad \Delta_k = \mu_k^* + \mu_k {r_1},
\end{equation}
isolating a fixed background and an $\mathcal{O}(r_1)$ part in the chemical potentials of the fugacities.
Then, depending on the specific background chosen for the real masses either all the massive KK modes for the matter and gauge fields become massive and decouple or some zero modes remain, generating a 3d gauge theory after the reduction.
Concretely, let us consider the case of a chiral multiplet charged under a $\U(1)$ gauge group, with R-charge $R$ and real mass $m$ under some $\U(1)$ flavor charge. Let us define some background $\mu^*$ for $\mu = m + \omega R$ and expand the gauge holonomy near some holonomy $\sigma^*$. The partition function of a chiral multiplet on $S^1 \times S_b^3$ is just an elliptic Gamma function. Let us denote $k = \sigma^* + \mu^*$ and $x r_1 = (\sigma + \mu)r_1$. Then, we have:
\begin{equation}
    \tilde \Gamma_e(k + x r_1) =  e^{\pi i Q(k + x r_1)} \prod_{n= -\infty}^{\infty} e^{- \mathrm{sign}(n) \frac{\pi i }{2\omega_1 \omega_2} \left( \left(\frac{k + n}{r_1}  + x -\omega \right)^2 - \frac{\omega_1 ^2 + \omega_2 ^ 2}{12} \right)}
    \Gamma_h \left(\frac{k + n}{r_1}  + x \right).
\end{equation}
Before proceeding with the reduction we will put ourselves in the region $0\leq k \leq 1$ using the periodicity of the elliptic Gamma functions $\tilde \Gamma_e(u+1;\tau,\sigma) = \tilde \Gamma_e(u;\tau,\sigma)$ so to avoid introducing periodic Bernoulli polynomials.
Now, depending on the choice of the background the whole KK tower can either lift or a zero mode can survive

\begin{equation}
    \label{eq:el_gamma_double}
    \Gamma_e(k + x r_1) \underset{r_1 \to 0}{\sim}
    \begin{cases}
        e^{-\frac{i \pi (x - \omega) }{ 6 r_1 \omega_1\omega_2}} \; \Gamma_h(x), \quad \quad & k \in \mathbb{Z} \\
        e^{i \pi Q(k + x r_1)}, & k \notin \mathbb{Z} \quad \land \quad 0 < k < 1,
    \end{cases}
\end{equation}

\begin{eqnarray}
Q(k + xr_1) =&& - \frac{1}{\omega_1 \omega_2} \Bigg( \frac{ k (2k - 1)(k - 1)}{6 r_1^2} + \frac{(x - \omega)\left( 6k (k - 1) + 1 \right)}{6r_1} + 
\nonumber \\ 
&& + \frac{(2k - 1)(6x^2 + \omega_1^2 + \omega_2^2 + 3 \omega_1\omega_2 - 12x\omega )}{12} \Bigg) + \mathcal{O}(r_1).
\end{eqnarray}

When the 3d reduction with the prescription of \cref{eq:double_scal_pr} is performed on a 4d gauge theory, the original gauge group is generically broken to a product of subgroups depending on the vacuum configuration. In principle, the reduction may generate CS and FI terms for general gauge group and generic vacuum, through the decoupling of the massive KK modes in \cref{eq:el_gamma_double}. Depending on the details of the theory the generation of some of these terms can be obstructed. In addition, this can be prevented by focusing on specific points for the background of the real masses and the vacua. For instance, notice that in the mirror point $k = 1/2$, $Q(1/2 + x r_1)$ simplifies and obstructs the generation of any CS term.

In such cases the original theory factorizes at the level of the partition function into a product of theories on the circle at finite small $r_1$, interacting only through monopole superpotential terms. The monopole superpotential is implemented through the balancing conditions constraints, which are crucial to preserve the duality at the level of the partition functions.

The matter content of such theories depends crucially on the choice of background for the real masses, namely a tuning of the background such that $\mu^* + \sigma^* \in \mathbb{Z}$ is required to preserve some KK zero modes in \cref{eq:double_scal_pr} and reduce to a 3d gauge theory with non-empty matter content.

After the choice of some background $ \mu = (\mu_1^*,\, \mu_2^*,\, \dots)$, which splits the flavor group $F \to F_1 \times F_2 \times \dots $, and a vacuum $\langle \sigma \rangle = (\sigma_1^*,\,\sigma_2^*,\, \dots)$, for which the original gauge group splits into $G \to G_1 \times G_2 \times \dots $, the index reduces as
\begin{equation}
    \mathcal{I}_{4d} \sim \, \prod_{i} \mathcal{ Z}^{(i)} \; \prod_{i,\,j} e^{\Phi_{ij}},
\end{equation}
where each $\mathcal{Z}^{(i)}$ is the partition function  defined in (\ref{eq:3dparf}) of the i-th gauge theory, with gauge group $G_i$, with its matter content and flavor symmetry identified by $F_a$, obtained by the reduction  
\begin{eqnarray}
    \mathcal{I}_{4d}^{(i)} && =\frac{(p;p)_\infty^{\mathrm{rk}G_i} (q;q)_\infty ^{\mathrm{rk}G_i}}{|W_i|} 
    \int \prod_{i=1}^{\mathrm{rk}G_i} \differential u_i \frac{1}{\prod_{\alpha \in \Delta_+}  \Gamma_e( z ^{\alpha(u)}) \Gamma_e( z ^{-\alpha(u)})}\dots \underset{r_1 \to 0}{\sim}
    \nonumber \\
    && \sim  \frac{e ^ { - \frac{i \pi \omega |G_i|}{6 r_1\omega_1 \omega_2}}} {|W_i|} 
    \int \prod_{i=1}^{\mathrm{rk}G_i} \frac{\differential \sigma_i}{\sqrt{- \omega_1\omega_2}} \frac{1}{\prod_{\alpha \in \Delta_+}  \Gamma_h( \alpha(\sigma) )\Gamma_h( - \alpha(\sigma) )}\dots,
\end{eqnarray}
where the dots stand for the matter content. The asymptotic expansion of the Pochhammer symbols can be derived from the Dedekind Eta function and its modular properties
\begin{equation}
    (p;p)_\infty = e ^{-\frac{ i\pi \tau}{12}}\eta(\tau) \underset{r_1 \to 0}{\sim} - \frac{\pi i}{12}\left(\tau + \frac{1}{\tau}\right) - \frac{1}{2}\log(-i\tau).
\end{equation}
The term $e^{\Phi_{ij}}$ represents the mixing between the holonomies and real masses for all the matter and vector KK modes in the \textit{i-th} and the \textit{j-th} gauge model on the circle, for which the corresponding hyperbolic Gamma functions are lifted
\begin{equation}
    e^{\Phi_{ij}}  = e ^ { i \pi \sum_{\mathrm{fields}} Q \left( k + x r_1 \right) }.
\end{equation}
The divergent terms $e^{\Phi_{ij}}$ arising from the lifting of the whole KK tower in this prescription for the reduction reveal to be crucial in preserving dualities, as they need to be identical in both dual phases. Identifying the dual split of the gauge group is crucial in order to properly recover those terms. 
In the next section we will use this prescription to recover 3d dualities from 4d theories in various examples. We will start discussing unitary and symplectic SQCD. Then, we will apply the prescription to the orthogonal case. In all the examples studied we will always consider a splitting of the gauge and flavor groups into a product of two subgroups for simplicity.

%
%
%
\section{Unitary and symplectic cases}
\label{sec:SUUSp}
%
%
%

\subsection*{Double scaling for $\U(N)$ Seiberg duality}

Here we start discussing the reduction of ordinary $\U(N)$ Seiberg duality with $F > N$ flavors on a circle.
The matter content and charges of the electric theory are: 
\begin{table}[htb] 
    \centering
    \begin{tabular}{cccccc}
        \toprule
          & $\U(N)$ & $\SU(F)_1$ & $\SU(F)_2$ & $\U(1)_B$ & $\U(1)_R$ \\
        \midrule
        $\mathcal{Q}$ & \scriptsize{\yng(1)} &  \scriptsize{\yng(1)} & $\mathbf{1}$ & 1 &  $ \frac{F-N}{F} $ \\
        $\tilde {\mathcal{Q}}$ & \textoverline{\scriptsize{\yng(1)}} & $\mathbf{1}$ & \textoverline{\scriptsize{\yng(1)}} & -1 & $ \frac{F-N}{F} $ \\
        \bottomrule
    \end{tabular} 
    \caption{Matter content and charges of 4d $\U(N)$ SQCD.}
        \label{matterSQCD}
\end{table}
while the matter content and charges of the dual magnetic theory are 
\begin{table}[htb] 
    \centering
    \begin{tabular}{cccccc}
        \toprule
          & $\U(F-N)$ & $\SU(F)_1$ & $\SU(F)_2$ & $\U(1)_B$ & $\U(1)_R$ \\
        \midrule
        $q$ & \scriptsize{\yng(1)} &  \textoverline{\scriptsize{\yng(1)}} & $\mathbf{1}$ & $\frac{N}{F - N}$ &  $ \frac{N}{F} $ \\
        $\tilde {q}$ & \textoverline{\scriptsize{\yng(1)}} & $\mathbf{1}$ & \scriptsize{\yng(1)} & -$\frac{N}{F - N}$ & $ \frac{N}{F} $ \\
        $M$ & $\mathbf{1}$ & \scriptsize{\yng(1)} &\textoverline{\scriptsize{\yng(1)}} & 0 & $2\frac{F-N}{F}$\\
        \bottomrule
    \end{tabular} 
    \caption{Matter content and charges of the magnetic theory $\U(\tilde N)$ SQCD.}
      \label{matterSQCDdual}
\end{table}

After weakly gauging the flavor symmetry we choose a configuration with $F-h$ flavors at the origin and $h$
flavors at the mirror point. The gauge holonomies split in $N-\ell$ at the origin and $\ell$  and  at the mirror point.
We  require $F-h \geq N-\ell$ and $h \geq \ell$ in order to have a stable configuration.

At finite $r_1$ we have an IR effective 3d dynamics, where  the fundamental monopoles $X_{1}^{\pm}$ and $X_{2}^{\pm}$ of $\U(N-\ell)$ and $\U(\ell)$ interact through a superpotential 
\begin{equation}
W = X_1^{+} X_2^{-} + X_1^{-} X_2^{+}. 
\end{equation}
Observe that this superpotential breaks both one combination of the two topological symmetries and one combination of the two axial symmetries of the 
two gauge theories.
The signs $\pm$ in $X_{1,2}^{\pm}$ then refer to the surviving topological symmetry.
In the dual model the corresponding vacuum preserving the 4d duality requires $F-h - N+\ell$ holonomies at the origin and $h-\ell$ holonomies 
at the mirror point.
The dual superpotential is 
\begin{equation}
\label{dualWU}
W = \sum_{I=1,2} M_I q_I \tilde q_I + \tilde X_1^{+} \tilde X_2^{-} + \tilde X_1^{-} \tilde X_2^{+}, 
\end{equation}
where the index $I$ labels the two dual gauge groups, $\U(F-h - N+\ell)$ at the origin and  $\U(h-\ell)$  one at the mirror point, 
while  the dual fundamental monopoles are indicated with a tilde.
Again a combination of the two topological symmetries and a combination of the two axial symmetries are broken by the superpotential (\ref{dualWU}).

Locally we can regard the $\U(h-\ell)$ theory as an effective 3d gauge theory at large $1/2r_1$ distance in the Coulomb branch with respect to the $\U(F-h - N+\ell)$ gauge theory.
For this reason  we   study the $r_1 \rightarrow 0$ limit, after performing a local duality on the $\U(h-\ell)$ SQCD.
We locally dualize $\U(h-\ell)$ by
using pure 3d Aharony duality, and we obtain a $\U(\ell)$ gauge theory with superpotential 
\begin{equation}
W=M_2 N_2+N_2 q_2 \tilde q_2 +   X_2^{-} \tilde X_2^{+}  + X_2^{+}  \tilde X_2^{+} \,.
\end{equation}
In this case the fields $ \tilde X_2^{\pm} $ correspond to singlets of the $\U(\ell)$  gauge theory, and we denoted the fundamental monopoles of this last
as $ X_2^{\pm}$ as above.
The superpotential, after integrating out the massive fields, becomes
\begin{equation}
W = M_1 q_1 \tilde q_1 +   X_2^{-} \tilde X_2^{+}  + X_2^{+}  \tilde X_2^{+} +  \tilde X_1^{+} \tilde X_2^{-} + \tilde X_1^{-} \tilde X_2^{+} .
\end{equation}
The role of the singlets $ \tilde X_2^{\pm} $ is to identify the monopoles of the $\U(\ell)$  sector with the ones of the $\U(F-h - N+\ell)$ sector.
Then we decouple from the electric and from the magnetic theory the sectors $\U(\ell)$ in the $r_1 \rightarrow 0$ limit, observing that while in the electric theory this removes the 
monopole superpotential in the dual theory there is still an effect due to the presence of the singlets $\tilde X_2^{\pm}$. Such singlets have the same quantum numbers of the monopoles $X_1^{\pm}$, and they set the monopoles $\tilde X_1^{\pm}$ to zero in the chiral ring. We end up with electric $\U(N-\ell)$ SQCD with $F - h$ flavors and vanishing superpotential, dual to magnetic $\U(F-h - N+\ell)$ SQCD with $F-h$ flavors and superpotential
\begin{equation}
W \propto M_1 q_1 \tilde q_1 +   X_1^{+}  \tilde X_1^{-} +  X_1^{-} \tilde X_1^{+}, 
\end{equation}
corresponding to Aharony duality.
In this derivation we have used, locally, the duality itself; however one way to ``circumvent'' this procedure consists in considering $h=1$ and $\ell=0$ such that the dual theory corresponds to SQED and where one can locally use mirror symmetry instead of Aharony duality.
Actually such an approach is not different from the one we took here because Aharony duality reduces to SQED/XYZ duality for $\U(1)$ with one flavor.
   
Let's see then how this entire process is formulated on the reduction of $\mathcal{I}_{4d}$ to $\mathcal{Z}_{3d}$.
Defining $\Delta_a$ and $\tilde \Delta_a$ the 4d fugacities of the fundamentals and the anti-fundamentals charged under the global  $\SU(F)_1\times \SU(F)_2\times \U(1)_B\times \U(1)_R$ symmetry in Table \ref{matterSQCD}, the index for the electric theory is 
\begin{equation}
    \mathcal{I}_{\U(N)}^{el} = \frac{(p;p)_\infty^N (q;q)_\infty^N}{N!} 
    \int \prod_{i=1}^{N} \mathrm{d}u_i \frac{\prod \limits_{a=1}^{F}\prod \limits_{i=1}^{N} \Gamma_e(u_i + \Delta_a)\Gamma_e( - u_i + \tilde \Delta_a)} 
    {\prod\limits_{i \neq j}  \Gamma_e(u_i - u_j )},
\end{equation}
with the balancing condition
\begin{equation}
    \label{eq:BCU}
    \sum_{a=1}^{F} (\Delta_a+\tilde \Delta_a) =2r_1\omega(F-N).
\end{equation}

When considering the effective field theory on the circle the flavor fugacities in the superconformal index become the real scalars in the background vector 
multiplets of the effective 3d description. 
We then turn on a background for such real masses and consider the theories around a vacuum configuration for the integration variables identified with the gauge holonomies, associated, in the 3d effective description, to the real scalars in the $\U(N)$ and $\U(F-N)$ vector multiplets.
Such vacuum configuration corresponds to fix one set of these integration variables at the origin, corresponding on the field theory side to the origin of the Coulomb branch.
On the other hand  the second set is fixed at the mirror point on the circle, that, in the field theory interpretation corresponds to a point at $1/r_1$ distance in the Coulomb branch 
with enhanced gauge symmetry.

The original 4d theory is split, in this 3d effective picture,  into a product of gauge theories (see Figure \ref{fig:1} LHS), with breaking pattern and flavor content 
\begin{equation}
\U(N) \, w/ \, F\text{ flavors} \to \U(N - \ell) \,w/ \,F-h \text{ flavors} \otimes \U(\ell)\, w/  \,h \text{ flavors},
\end{equation}
labeled by two integers $\ell=1,\dots,\, N$ and $h=1,\dots,F$. From the point of view of the partition function we have two decoupled integrals over the two sets of 
integration variables $\sigma$ obtained by expanding the 4d gauge holonomies $u$ around the origin and the mirror point respectively. Furthermore, the 4d flavor fugacities $\Delta$ and $\tilde \Delta$ become the 3d real masses $\mu$ and $\nu$ according to
\begin{eqnarray}
    \label{eq:backU}
        &&\begin{cases}
        u_i = \sigma_i r_1 &\quad \quad i=1,\dots, N - \ell\\
        u_{N - \ell + i} = \frac{1}{2} + \sigma_{N - \ell + i} r_1 &\quad \quad i=1,\dots, \ell 
    \end{cases}
        \nonumber \\
    &&\begin{cases}
       \Delta_a = \mu_a r_1 &\quad \;\;\quad a=1,\dots, F - h\\
       \tilde \Delta_a = \nu_a r_1 &\quad \;\;\quad a=1,\dots, F - h\\
       \Delta_{F - h + a} = \frac{1}{2} + \mu_a r_1 &\quad \;\;\quad a= 1,\dots, h \\
       \tilde \Delta_{F - h + a} = - \frac{1}{2} + \nu_a r_1 &\quad \;\;\quad a=1,\dots, h
    \end{cases}
\end{eqnarray}
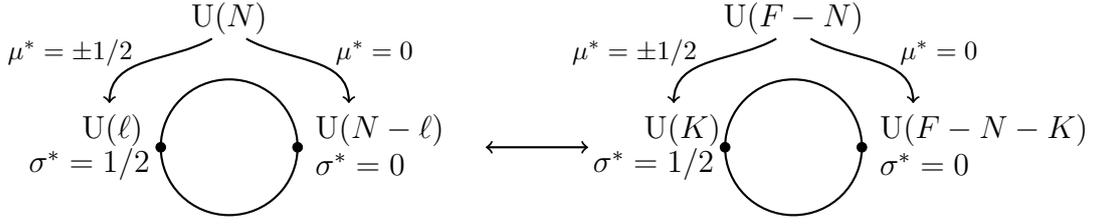
\begin{figure}[h!]
    \centering
        \begin{tikzpicture}[scale=.45]


          \draw node at (0,3.8) {$\U(N)$};
          \draw[thick] (0,0) circle(2cm);
          \filldraw[fill=black] (2,0) circle (4pt);

          \draw[thick,<-] (3.5,1.3) to[out=90, in=-30] (0.5,3.2);
          \draw[above,left] node at (-2.5,2.8) {\footnotesize$ \mu^* = \pm 1/2$};
          \draw[thick,<-] (-3.5,1.3) to[out=90, in=-150] (-0.5,3.2);
          \draw[above,right] node at (2.8,2.8) {\footnotesize$ \mu^* = 0$};

          \draw node[right] at (2.2,0.5) {${\U(N - \ell)}$};
          \draw node[right] at (2.2,-0.5) {$\sigma^* = 0$};

          \draw node[left] at (-2.2,0.5) {$\U(\ell)$};
          \draw node[left] at (-2,-0.5) {$\sigma^* = 1/2$};
          \filldraw[fill=black] (-2,0) circle (4pt);

          \draw[thick,<->] (7.5,0) to[out=0, in=180] (10.5,0);


          \draw node at (16.5,3.8) {$\U(F-N)$};
          \draw[thick] (16.5,0) circle(2cm);
          \filldraw[fill=black] (18.5,0) circle (4pt);
  
          \draw[thick,<-] (20,1.3) to[out=90, in=-30] (17,3.2);
          \draw[above,left] node at (14,2.8) {\footnotesize$ \mu^* = \pm 1/2$};
          \draw[thick,<-] (13,1.3) to[out=90, in=-150] (16,3.2);
          \draw[above,right] node at (19.3,2.8) {\footnotesize$ \mu^* = 0$};
  
          \draw node[right] at (18.7,0.5) {${\U(F - N - K)}$};
          \draw node[right] at (18.7,-0.5) {$\sigma^* = 0$};
  
          \draw node[left] at (14.7,0.5) {$\U(K)$};
          \draw node[left] at (14.5,-0.5) {$\sigma^* = 1/2$};
          \filldraw[fill=black] (14.5,0) circle (4pt);
        \end{tikzpicture}
     \caption{\footnotesize{Split  into a product of gauge theories in the 3d effective picture, both in the electric and in the magnetic unitary SQCD.}}
    \label{fig:1}
\end{figure}

The 4d index in this effective 3d picture can be obtained by considering the whole KK theory on $S^1_{r_1}$ on this non-trivial flavor background.  
The index in this case splits as 
\begin{equation}
    \label{eq:index_un}
    \mathcal{I}_{\U(N)}^{el} = \frac{(p;p)_\infty^N (q;q)_\infty^N}{N!} \int \prod_{i=1}^{N} \differential u_i {I}_{\U(N_1)}^{F_1}I_{\U(N_2)}^{F_2} I_{(1,\,2 )},
\end{equation}
where
\begin{equation}
    I_{\U(N_i)}^{F_i} =\frac{\prod \limits_{a=1}^{F_i}\prod \limits_{i=1}^{N_i} \Gamma_e(u_i + \Delta_a)\Gamma_e( - u_i + \tilde \Delta_a)} 
    {\prod\limits_{1\leq i \neq j \leq N_i}  \Gamma_e(u_i - u_j )}
\end{equation}
and upon using the periodicity under integral shifts of the elliptic gamma functions, we defined
\begin{eqnarray}
    I_{(1,\, 2 )} = && \prod_{a=1}^{F_1}\prod_{i = 1}^{N_2} \Gamma_e \left(1/2 + u_{N_1 + i} + \Delta_a \right) \Gamma_e \left(1/2 - u_{N_1 + i} + \tilde \Delta_a \right) \nonumber \\
    && \prod_{a=1}^{F_2}\prod_{i = 1}^{N_1} \Gamma_e \left(1/2 + u_{i} + \Delta_{F1 + a} \right) \Gamma_e \left(1/2 - u_i +   \tilde \Delta_{F_1 + a} \right)
    \nonumber \\
    && \left(\prod_{i=1}^{N_2}\prod_{j=1}^{N_1} \Gamma_e \left(1/2  + u_{N_1+i} - u_j\right) \prod_{i=1}^{N_1}\prod_{j=1}^{N_2} \Gamma_e \left(1/2  + u_{i} - u_{N_1 + j}\right) \right) ^{-1}.
\end{eqnarray}
Taking the  limit (\cref{eq:el_gamma_double}) in this background and employing \cref{eq:BCU} we get
 \begin{equation}
    I_{(1,\,2 )} \sim e^{\Phi_{1,2}^{el}},
 \end{equation}
 with, including also the $\mathcal{O}\left(\frac{1}{r_1}\right)$ contribution arising from the reduction of the Pochhammer symbols for later convenience,
 \begin{align}
    & \Phi_{1,2}^{el\;'} = \Phi_{1,2}^{el} + \frac{i \pi N}{6r_1 \omega_1\omega_2}
    = 
    \nonumber \\
    = & i \pi \left(\sum_{\varphi \in \mathcal{Q},\tilde {\mathcal{Q}}} Q (1/2 + \rho_{\varphi}(u) + \Delta_\varphi) - \sum_{V} Q (1/2 + \alpha(u))+ \frac{N}{6r_1 \omega_1\omega_2} \right) = 
    \nonumber \\
    = & \frac{i \pi}{12 r_1 \omega_1 \omega_2} 
     \Bigg( 2\omega ( 3\ell(2h - \ell) + N^2 - 3hN) +\, 3(N - 2\ell)\sum_{a = F - h + 1}^F (\mu_a + \nu_a)\Bigg).
 \end{align}
 Then, as discussed above,  the index (\ref{eq:index_un}) factorizes into a product of two decoupled 3d partition functions:
\begin{equation}
\label{eq:4d3dunKKmon}
    \mathcal{I}_{4d}^{el} \underset{r_1 \to 0}{\sim}e^{\Phi_{1,2}^{el\;'}} \mathcal{Z}_{\U(N-\ell)}^{N - h} \mathcal{Z}_{\U(\ell)}^{h}.
\end{equation}
Actually the 3d theories associated to the partition functions on the RHS of (\ref{eq:4d3dunKKmon}) interact through their fundamental monopoles as discussed above.
On the partition function side such interaction is reflected by the balancing condition
\begin{equation}
    \label{eq:BCmu}
    \sum_{a=1}^{F} (\mu_a+\nu_a) =2\omega(F-N),
\end{equation}
arising from (\ref{eq:BCU}).

Next we study the reduction of the 4d dual magnetic theory.
The index of this theory can be written from the charges in Table \ref{matterSQCDdual} and, by defining $\tilde N = F - N$, it corresponds to the 
integral 
\begin{eqnarray}
    \mathcal{I}_{\U(\tilde N)}^{mag} = &&\frac{(p;p)_\infty^ {\tilde N}  (q;q)_\infty^{\tilde N}} {\tilde N !}\prod_{a,b=1}^{F}\Gamma_e (\Delta_a + \tilde \Delta_b) 
    \nonumber \\
    &&\int \prod_{i=1}^{\tilde N} \mathrm{d}u_i \frac{\prod\limits_{a=1}^F \prod\limits_{i=1}^{\tilde N} \Gamma_e(u_i + r_1\omega - \Delta_a )\Gamma_e( - u_i + r_1\omega - \tilde \Delta_a)} 
    {\prod\limits_{1\leq i \neq j \leq \tilde N}  \Gamma_e(u_i - u_j )}.
\end{eqnarray}

We consider the  limit (\cref{eq:el_gamma_double}) with non-trivial background of the form (\ref{eq:backU}). At this level of the discussion we split the flavor symmetry as in the electric side, because it is a necessary condition to maintain the duality. On the other hand we split the gauge holonomies by keeping $F-N-K$ at the origin and $K$ at the mirror point (see Figure \ref{fig:1} RHS).
We arrive at
\begin{equation}
\label{eq:effdualZSQCD}
    \mathcal{I}_{4d}^{mag} \underset{r_1 \to 0}{\sim} e^{\Phi_{1,2}^{mag\;'}} \hat{\mathcal{Z}}_{\U(F - N - K )}^{F - h} \hat{\mathcal{Z}}_{\U(K)}^{h},
\end{equation}
where the notation $\hat{\mathcal{Z}}_{\U(N_i)}^{{F_i}}$ also includes the contribution from the meson and thus 
\begin{equation}
\hat{\mathcal{Z}}_{\U(N_i)}^{F_i} = \prod_{a,b = 1}^{F_i} \Gamma_h(\mu_a + \nu_b) \mathcal{Z}_{U(N_i)}^{F_i},
\end{equation}
while
\begin{equation}
    \Phi_{1,2}^{mag \; '} = \frac{i \pi}{12 r_1 \omega_1 \omega_2}   
    \left( 2\omega(3h^2 - 3K^2 + N  - 3hN) + 3(N - 2h + 2K)
   \!\!\! \sum_{a = F - h + 1}^F   \!\!\! (\mu_a + \nu_a) \right).
\end{equation}
In order to maintain the integral identity between (\ref{eq:4d3dunKKmon}) and (\ref{eq:effdualZSQCD})
the mapping of the vacua across the dual phases requires  $K = h - \ell$. This indeed reflects the 
result obtained from the field theory approach.

The relation $K = h - \ell$ is precisely the constraint that emerges by the duality requirement $\Phi_{(1,\,2 )}^{el \;'} = \Phi_{(1,\,2 )}^{mag \;'}$, and it can be read off directly from the requirement
\begin{equation}
\label{eq:constraintPhi}
    \Phi_{1,\,2 }^{mag \;'} - \Phi_{1,\,2 }^{el \;'} = \frac{i\pi(h - K - \ell)}{2 r_1\omega_1\omega_2}\left( \omega (h + K - \ell) - \sum_{a=F - h + 1}^{F}(\mu_a + \nu_a) \right) \overset{!}{=}0.
\end{equation}

Notice that $K = h - \ell$ implies $h > \ell$ and $h > K$, which are the vacuum stability condition on the number of flavors in both the dual theories. After imposing (\ref{eq:constraintPhi}) on the identity between 
(\ref{eq:4d3dunKKmon}) and (\ref{eq:effdualZSQCD})
 the effective 3d duality  on the partition function corresponds to the identity: 
\begin{eqnarray}
\label{eq:U(N)onS1}
&&\mathcal{Z}_{\U(N-\ell)} (\mu^{(1)},\nu^{(1)},\lambda) \mathcal{Z}_{\U(\ell)}(\mu^{(2)},\nu^{(2)},\lambda)
=
\prod_{a,b=1}^{F-h} \Gamma_h(\mu_a^{(1)} +\nu_b^{(2)})  \prod_{a,b=1}^{h} \Gamma_h(\mu_a^{(2)}+\nu_b^{(2)}) 
\nonumber \\
\times&&
\mathcal{Z}_{\U(F-h-N+\ell)}(\omega-\mu^{(1)},\omega-\nu^{(1)},-\lambda) \mathcal{Z}_{\U(h-\ell)}(\omega-\mu^{(2)},\omega-\nu^{(2)},-\lambda),
\end{eqnarray}
with the balancing condition (\ref{eq:BCmu}).
Observe that in this formula we have inserted back the explicit dependence of the partition functions from the flavor fugacities and from the topological 
symmetry.
We split the fugacity vectors $\mu$ and $\nu$ as $(\mu^{(1)},\mu^{(2)})$ and  $(\nu^{(1)},\nu^{(2)})$ respectively, corresponding to the split into $F-h$ and $h$ real masses
for both the fundamentals and anti-fundamentals. The parameter $\lambda$ labels the FI terms in the various gauge sector and their follow with the duality map.

The 3d limit is obtained by first applying locally the integral identity corresponding to $\U(N)$ Aharony duality on the last integral on the RHS of (\ref{eq:U(N)onS1}). Such identity 
corresponds in this setup to the relation
\begin{eqnarray}
\label{eq:Aha3dfromdua}
&&
 \mathcal{Z}_{\U(h-\ell)}(\omega-\mu^{(2)},\omega-\nu^{(2)},-\lambda)
 =
 \Gamma_h \left(\omega(\ell+1)+\frac{1}{2} \sum_{a=1}^{h}(
 \mu_a^{(2)} +\nu_a^{(2)} )\pm \frac{\lambda}{2} \right)
 \nonumber \\
\times &&
 \prod_{a,b=1}^{h} \Gamma_h(2\omega-\mu_a^{(2)}-\nu_b^{(2)}) 
\mathcal{Z}_{\U(\ell)}(\mu^{(2)},\nu^{(2)},\lambda).
\end{eqnarray}
Observe that plugging this relation into (\ref{eq:U(N)onS1}) corresponds in the field theory approach
to the local use of the pure 3d duality in the effective description. Then
we can rearrange the contribution of the singlets in the first term on the RHS of (\ref{eq:Aha3dfromdua})
 using the balancing condition (\ref{eq:BCmu}).
 
 At this point the final step corresponds to get rid of the common integrals $\mathcal{Z}_{\U(\ell)}(\mu^{(2)},\nu^{(2)},\lambda)$.
 This indeed corresponds to consider the pure 3d limit where the physics at large distance in the Coulomb branch is effectively decoupled.
 The monopoles of the electric theory acting as singlets in the dual model are not lift by the limit, and they appear in the final identity, that is 
\begin{eqnarray}
\label{eq:U(N)3d}
&&
\mathcal{Z}_{\U(N-\ell)}(\mu^{(1)},\nu^{(1)},\lambda) 
=
\Gamma_h \left(\omega(F-h-N+1)-\frac{1}{2} \sum_{a=1}^{F-h}(
 \mu_a^{(1)} +\nu_a^{(1)} )\pm \frac{\lambda}{2} \right)
 \nonumber \\
&&
\prod_{a,b=1}^{F-h} \Gamma_h(\mu_a^{(1)} +\nu_b^{(1)}) 
\mathcal{Z}_{\U(F-h-N+\ell)}(\omega-\mu^{(1)},\omega-\nu^{(1)},-\lambda), 
\nonumber \\
\end{eqnarray}
which holds without any balancing condition. This final identity is the partition function realization of 3d Aharony duality, obtained here
 from the double scaling limit of the identity between the supersymmetric indices of $\U(N)$ 4d Seiberg duality.

\subsection*{Double scaling for $\USp(2N)$ Intriligator--Pouliot duality}

A similar study can be pursued for the case of Intriligator--Pouliot duality involving a $\USp(2N)$ gauge group with $2F$ fundamentals. 
The matter content and charges of the electric theory are: 
\begin{table}[htb] 
    \centering
    \begin{tabular}{ccccc}
        \toprule
          & $\USp(2N)$ & $\SU(2F)$ & $\U(1)_R$ \\
        \midrule
        $\mathcal{Q}$ & \scriptsize{\yng(1)} &  \scriptsize{\yng(1)} &  $ \frac{F-N-1}{F} $\\
        \bottomrule
    \end{tabular} 
    \caption{Matter content and charges of $\USp(2N)$ SQCD.}
    \label{eleUSp}
\end{table}
\\
The matter content and the charges of the magnetic theory are:
\begin{table}[htb] 
    \centering
    \begin{tabular}{ccccc}
        \toprule
          & $\USp(2F - 2N - 4)$ & $\SU(2F)$ & $\U(1)_R$ \\
        \midrule
        $q$ & \scriptsize{\yng(1)} &  \textoverline{\scriptsize{\yng(1)}} &  $ \frac{N + 1}{F} $\\
        $M$ & $\mathbf{1}$ &  \scriptsize{\yng(1,1)} &  $ 2\frac{F-N-1}{F} $\\
        \bottomrule
    \end{tabular} 
    \caption{Matter content and charges of magnetic $\USp(2F - 2N - 4)$ SQCD.}
    \label{magUSp}
\end{table}

Splitting again $F$ in $F-h$ and $h$, 
 $N$ in $N-\ell$ and $\ell$ in the dual phase, we need to split the gauge holonomies as $F - h - N+\ell - 1$ and  $h - \ell - 1$.
The superpotential of the electric theory involves the fundamental  monopoles $Y_1$ and $Y_2$ of $\USp(2(N-\ell))$ and $\USp(2\ell)$ respectively, and it is $W \propto Y_1 Y_2$. Analogously, the dual superpotential is 
 \begin{equation}
W = \sum_{I=1,2} M_I q_I \cdot q_I + \tilde  Y_1 \tilde Y_2.  
\end{equation}
The 3d limit is found by first applying Aharony duality on $\USp(2(h-\ell-1))$ finding  $\USp(2 \ell)$.
The $r_1\rightarrow 0$ limit gives then the expected pure 3d duality, where the fundamental monopole of $\USp(2(N-\ell))$ 
acts as a singlet in the $\USp(2(F-h-N+\ell-1))$ gauge theory, and it
is identified with the singlet arising by the local duality on $\USp(2(h-\ell))$ described above.
\\
Next, we will study such reduction of the duality with the  double scaling on the flavor fugacities and on the radius $r_1$ in the reduction of 
the 4d index to the 3d partition function.

The index of the electric theory can be written from the charges in Table \ref{eleUSp}
\begin{equation}
    \mathcal{I}_{\USp(2N)}^{el} = \frac{(p;p)_\infty^N (q;q)_\infty^N}{2^N N!} 
    \int \prod_{i=1}^{N} \mathrm{d}u_i \frac{\prod\limits_{a = 1}^{F} \prod\limits_{i = 1}^N \Gamma_e(\pm u_i + \Delta_a)} 
    {\prod\limits_{1 \leq i<j \leq N}  \Gamma_e(\pm u_i \pm u_j)\prod_i \Gamma_e(\pm 2 u_i)},
\end{equation}
together with the balancing condition
\begin{equation}
\label{eq:BCUSp}
\sum_{a=1}^{2F} \Delta_a = 2r_1 \omega(F - N - 1).
\end{equation}
Taking the limit (\ref{eq:el_gamma_double}) in this case requires turning on an appropriate background for the real masses and the real scalars in the vector multiplet in the effective 3d description, in order to preserve the parity of the flavors in the effective 3d theories emerging from the decoupling limit of the massive KK tower, and the balancing condition
\begin{equation}
    \label{eq:bcUSp3d}
    \sum_{a = 1}^{2F} \mu_a = 2\omega (F - N - 1).
\end{equation}
Thus, we take the limit (\ref{eq:el_gamma_double}) on the following configuration of 4d gauge holonomies and flavor fugacities:
\begin{eqnarray}
    \label{eq:backUSp}
    &&\begin{cases}
       \Delta_{2a\phantom{+1}} = \,\,\,\,\, \mu_{2a} r_1 -\frac{1}{2} &\quad \;\;\quad a=1,\dots, F - h\\
       \Delta_{2a - 1} =  \mu_{2a - 1}  r_1 +\frac{1}{2} &\quad \;\;\quad a= 1,\dots, F - h \\
       \Delta_{2F - 2h + a} = \mu_{2F - 2h + a} r_1 &\quad \;\;\quad a=1,\dots, 2h\\
    \end{cases}
    \nonumber \\
    &&\begin{cases}
        u_i = \sigma_i r_1 &\quad \quad i=1,\dots, N - \ell\\
        u_{ N - \ell + i} = \frac{1}{2} + \sigma_{ N - \ell + i} r_1 &\quad \quad i=1,\dots, \ell 
    \end{cases}
\end{eqnarray}
The 4d theory in the effective 3d picture splits accordingly:
 \begin{equation}
    \USp(2N)_{2F} \, w/ 2F \,\text{fund.} \to \USp(2N - 2\ell)\, w/ 2F-2h  \,\text{fund.} \times \USp(2\ell) \, w/ 2h \,\text{fund.} 
 \end{equation}
Then, employing \cref{eq:el_gamma_double}, the index reduces to
\begin{equation}
    \mathcal{I}_{4d}^{el} \underset{r_1 \to 0}{\sim}e^{\Phi_{1,2}^{el\;'}} \mathcal{Z}_{\USp(2N - 2\ell)}^{2F - 2h}\mathcal{Z}_{\USp(2\ell)}^{2h},
\end{equation}
with
\begin{equation}
    \Phi_{1,\,2 }^{el \;'} = \frac{i \pi}{12r_1\omega_1\omega_2}\left((6\ell(1-2h + \ell) - N(3-6h + 2N))2\omega + 6(N - 2\ell)\sum_{a = 2F - 2h + 1}^{2F}\mu_a \right).
\end{equation}

The index  of the electric theory can be written from the charges in Table \ref{magUSp} and, defining $\tilde N = F - N - 2$, it becomes
\begin{equation}
    \mathcal{I}_{\USp(2N)}^{mag} = \frac{(p;p)_\infty^{\tilde N} (q;q)_\infty^{\tilde N}}{2^{\tilde N} \tilde N!} \prod_{1\leq a<b\leq F}\Gamma_e (\Delta_a + \Delta_b)
    \int \prod_{i=1}^{\tilde N} \mathrm{d}u_i \frac{\prod_{a,i} \Gamma_e(\pm u_i + r_1\omega - \Delta_a)} 
    {\prod_{i,j}  \Gamma_e(\pm u_i \pm u_j)\Gamma_e(\pm 2 u_i)}.
\end{equation}
Reducing the theory on the configuration (\ref{eq:backUSp}), splitting the flavors as in the electric side and the gauge as $\tilde N -K$ and $K$, we get
\begin{equation}
    \mathcal{I}_{4d}^{mag} \underset{r_1 \to 0}{\sim}e^{\Phi_{1,2}^{mag\;'}} \hat{\mathcal{Z}}_{\USp(2 \tilde N - 2K)}^{2F - 2h}
    \hat{\mathcal{Z}}_{\USp(2K)}^{2h},
\end{equation}
with
\begin{align}
    \Phi_{1,\,2 }^{mag \;'} = \frac{i \pi}{12r_1\omega_1\omega_2} \Bigg( & (6h^2 - 6K(1 + K) -6h(1 + N) + N(3 + 2N))2\omega +
    \nonumber \\
    & + 6(2-2h +2K + N)\sum_{a = 2F - 2h + 1}^{2F}\mu_a \Bigg)
\end{align}
and with the relation
\begin{equation}
 \hat{\mathcal{Z}}_{\USp(2 N_i)}^{2F_i}
  = \prod_{1\leq a<b\leq F_i} \Gamma_h(\mu_a + \mu_b) \mathcal{Z}_{\USp(2 N_i)}^{2F_i}.
\end{equation}
Again we impose  
\begin{equation}
    \Phi_{1,\,2 }^{mag \;'} -  \Phi_{1,\,2}^{el \;'} = \frac{i\pi(h - K - \ell - 1)\left((h + K - \ell)\omega - \sum_{a = 2F - 2h + 1}^{2F} \mu_a \right)}{r_1\omega_1\omega_2} \overset{!}{=} 0,
\end{equation}
we identify the matching between dual vacua $K = h - \ell - 1$, for any $\ell = 1,\dots,N$ and $h = 1,\dots,F$. For the extremal cases $h = 0, F$ and $\ell = 0, N$ we get $K = 0$ and $K = F - N - 2$ respectively.

The effective 3d duality  on the partition function corresponds to the identity
\begin{eqnarray}
\label{USp(2N)onS1}
\mathcal{Z}_{\USp(2(N-\ell))}(\mu^{(1)}) \!\!\!\!&&\!\! \!\! \mathcal{Z}_{\USp(2\ell)}(\mu^{(2)})
=
\prod_{1\leq a<b\leq 2F- 2h} \Gamma_h(\mu_a^{(1)} +\mu_b^{(1)})  \prod_{1\leq a<b\leq 2h} \Gamma_h(\mu_a^{(2)}+\mu_b^{(2)}) 
\nonumber \\
&\times&
\mathcal{Z}_{\USp(2(F-h-N+\ell-1))}(\omega-\mu^{(1)}) \mathcal{Z}_{\USp(2(h-\ell-1))}(\omega-\mu^{(2)}),
\nonumber \\
\end{eqnarray}
with the balancing conditions (\ref{eq:bcUSp3d}). 
Again we have inserted back the explicit dependence of the partition functions from the flavor fugacities, and we have split
 the fugacity vectors $\mu$  as $(\mu^{(1)},\mu^{(2)})$, corresponding to the split into $2F-2h$ and $2h$ real masses
for  the fundamentals.

The 3d limit is obtained by first applying locally the integral identity corresponding to $\USp(2N)$ Aharony duality on the last integral on the RHS of (\ref{USp(2N)onS1}). Such identity corresponds in this setup to the relation
\begin{eqnarray}
\label{Aha3dUSpfromdua}
&&
\mathcal{Z}_{\USp(2(h-\ell-1))}(\omega-\mu^{(2)})
 =
 \Gamma_h \left(2\omega(\ell - h)+\sum_{a=1}^{2h}
 \mu_a^{(2)} \right)
 \nonumber \\
&&
 \prod_{1\leq a<b\leq 2h} \Gamma_h(2\omega-\mu_a^{(2)}-\mu_b^{(2)}) 
\mathcal{Z}_{\USp(2 \ell)}(\mu^{(2)}).
\end{eqnarray}

Observe that plugging this relation into (\ref{USp(2N)onS1}) corresponds in the field theory approach
to the local use of the pure 3d duality in the effective description. Then,
we can rearrange the contribution of the singlets in the first term on the RHS of (\ref{Aha3dUSpfromdua})
 using the balancing condition (\ref{eq:bcUSp3d}).
 
 At this point the final step corresponds to get rid of the common integrals $\mathcal{Z}_{\USp(2 \ell)}(\mu^{(2)})$.
 This indeed corresponds to consider the pure 3d limit where the physics at large distance in the Coulomb branch is effectively decoupled.
 The monopoles of the electric theory acting as singlets in the dual model are not lifted by the limit, and they appear in the final identity, that is 
\begin{eqnarray}
\label{USp(N)3d}
&&
\mathcal{Z}_{\USp(2(N-\ell))}(\mu^{(1)}) 
=
 \Gamma_h \left(2\omega(F - h - N+\ell - 1)- \sum_{a=1}^{2F-2h}
 \mu_a^{(1)} \right)
\nonumber \\
&&\prod_{1\leq a<b\leq 2F- 2h} \Gamma_h(\mu_a^{(1)} +\mu_b^{(2)}) 
\mathcal{Z}_{\USp(2(F-h-N+\ell-1))}(\omega-\mu^{(1)}),
\nonumber \\
\end{eqnarray}
which holds without any balancing condition. This final identity is the partition function realization of 3d Aharony duality, obtained here
 from the double scaling limit of the identity between the supersymmetric indices of 4d Intriligator--Pouliot duality.

%
%
%
\section{Orthogonal  case}
\label{sec:SO}
%
%
%

In this section we study the reduction of the 4d duality for orthogonal gauge groups with vectors originally studied in \cite{Intriligator:1995id}.
In this case the matter content does not break completely the center of the gauge group and various dualities are possible when considering 
type $B$ and $D$ gauge algebras \cite{Aharony:2013hda}.
Here we will mostly focus on the $\SO(N)_+$ case with $F$ vectors, where the dual theory is $\SO(F-N+4)_+$ with $F$ dual vectors interacting 
with a symmetric meson.
The 3d reduction of this theory from the field theory side has been deeply investigated in \cite{Aharony:2013kma}. The global properties here
play again a non-trivial role and various options are possible, involving $\mathrm{O}(N)_\pm$, $\mathrm{Spin}(N)$, $\SO(N)$ and $\mathrm{Pin}(N)$ gauge groups.
Such options are related to the possibility of (combined) discrete gaugings with respect to the $\mathbb{Z}_2$ charge conjugation and the $\mathbb{Z}_2$ magnetic symmetry.
Such gaugings have here dramatic consequences on the dualities because the matter content itself can be modified by such gaugings.
These possibilities have been summarized in \cite{Aharony:2013kma}, generalizing the results of 
\cite{Hwang:2011ht,Kapustin:2011gh,Aharony:2011ci,Benini:2011mf}, holding for the $\mathrm{O}(N)_+$ case\footnote{See also \cite{Benvenuti:2021nwt} for a deep analysis of the global properties for such dualities in the confining limit.}.
Even if the  identities matching the three sphere partition functions have been obtained for the $\mathrm{O}(N)_+$ dualities 
they imply the same identities for the other pure 3d dualities discussed in  \cite{Aharony:2013kma}, because the only difference
stays in the volume of the gauge groups, that is at most modified by a factor of 2.
In this section our goal consists then of deriving such 3d identities from the double scaling limit designed above, starting from the 4d identity for the duality of \cite{Intriligator:1995id}.
This last was originally derived in \cite{Dolan:2008qi}.
For these reasons we will not specify the global properties of the 4d theory, and we will mostly refer to the 4d $\SO(N)_+$ duality.

The charges and field content for $\SO (2N + \varepsilon)$, $\varepsilon = 0,1 $ SQCD with $F$ $\SU(F)$ fundamental vectors are
\begin{table}[htb] 
    \centering
    \begin{tabular}{cccc}
        \toprule
          & $\SO(2N + \varepsilon)$ & $\SU(F)$ & $\U(1)_R$ \\
        \midrule
        $Q$ & \scriptsize{\yng(1)} & \scriptsize{\yng(1)} &  $ \frac{F - N + 2}{F} $ \\
        \bottomrule
    \end{tabular} 
    \caption{Matter content and charges of orthogonal SQCD}
    \label{eleSOSQCD}
\end{table}
\\
We can write the index for $SO(2N + \varepsilon)$ SQCD by using the charges in Table \ref{eleSOSQCD} as
\begin{equation}
    I_\varepsilon^{el} = \frac{(p;p)_\infty^{N} (q;q)_\infty^{N}}{2^{N - (1 - \varepsilon)} N!} 
    \prod_{a=1}^{F} \Gamma_e( \Delta_a )^\varepsilon  
    \int \prod_{i=1}^{N} \mathrm{d}u_i \frac{\prod\limits_{a=1}^{F}\prod\limits_{i=1}^{N} \Gamma_e(\pm u_i + \Delta_a)} 
    {\prod\limits_{1\leq i < j \leq N}  \Gamma_e( \pm u_i \pm u_j )\prod\limits_{i=1}^{N} \Gamma_e( \pm u_i )^\varepsilon},
\end{equation}
with balancing condition
\begin{eqnarray}
    \label{BCO}
    \sum_{a=1}^{F} \Delta_a
    =
    r_1 \omega(F - 2N -\varepsilon + 2).
\end{eqnarray}
In the magnetic phase the field content and the charges are
\begin{table}[H] 
    \centering
    \begin{tabular}{cccc}
        \toprule
          & $\SO(F - 2N + 4 - \varepsilon)$ & $\SU(F)$ & $\U(1)_R$ \\
        \midrule
        $q$ & \scriptsize{\yng(1)} & \textoverline{\scriptsize{\yng(1)}} &  $ \frac{N-2}{F} $ \\
        $M$ & $\mathbf{1}$ & \scriptsize{\yng(2)} &  $ 2\frac{F-N+2}{F} $ \\
        \bottomrule
    \end{tabular} 
    \caption{Matter content and charges of the magnetic orthogonal SQCD.}
    \label{magSOSQCD}
\end{table}

The parity of the magnetic theory depends on the parity of the electric gauge group and the number of flavors, $F \equiv 2f + \varepsilon_f$, via the combination $\varepsilon_m = \varepsilon_f - \varepsilon$.
We can write the magnetic index  by using the charges in Table \ref{magSOSQCD} as
\begin{eqnarray}
    I_{\varepsilon_m} ^{mag} = \frac{(p;p)_\infty^{\tilde N} (q;q)_\infty^{\tilde N}}{2^{\tilde N - (1 - {\varepsilon_m})} \tilde N!} 
    \prod_{1\leq a \leq b \leq F} \Gamma_e(\Delta_a + \Delta_b) \prod_{a=1}^{F} \Gamma_e(r_1\omega - \Delta_a )^{\varepsilon_m}
    \nonumber \\
    \int \prod_{i=1}^{\tilde N} \mathrm{d}u_i \frac{\prod \limits_{a = 1}^{F} \prod \limits_{i=1}^{\tilde N} \Gamma_e(\pm u_i + r_1\omega - \Delta_a)} 
    {\prod \limits_{1\leq i < j \leq \tilde N}  \Gamma_e( \pm u_i \pm u_j ) \prod \limits_{i = 1}^{\tilde N} \Gamma_e( \pm u_i )^{\varepsilon_m}}.
\end{eqnarray}
We distinguish four cases depending on the parity of the electric and the magnetic theory respectively:
\begin{table}[h!] 
    \centering
    \begin{tabular}{ccccr}
        \toprule
         $\SO(2N + \varepsilon)$ & $\SO(2 \tilde N + |\varepsilon_m|)$ & $\varepsilon$ & $\varepsilon_f$& $\varepsilon_m$ \\
        \midrule
        $ D_N $ & $ D_{\tilde N} $ & 0 &  $ 0 $ &$0$ \\
        $ B_N $ & $ B_{\tilde N} $ & 1 &  $ 0 $ &$-1$ \\
        $ D_N $ & $ B_{\tilde N} $ & 0 &  $ 1 $ &$1$ \\
        $ B_N $ & $ D_{\tilde N} $ & 1 &  $ 1 $ &$0$ \\
        \bottomrule
    \end{tabular} 
    \caption{Electric and magnetic gauge groups for even and odd flavors $\varepsilon_f = 0,\, 1$. The gauge groups are schematically labeled with the name of the Dynkin diagram they belong to: $D_N \to \SO(2N)$ and $B_N \to \SO(2N + 1)$. }
    \label{tab:so}
\end{table}

Turning on a background for the real masses, we expand the theory around a vacuum configuration defined by two packages of holonomies, one at the origin of the Coulomb branch and one at the mirror point on the circle respectively. The original theory is split accordingly into a product of gauge theories, with breaking pattern
\begin{eqnarray}
\SO(2N + \varepsilon) \, w/ \,F \text{  vec.} \to \SO(2N + \varepsilon - 2\ell)\, w/ \,F-2h \text{  vec.}   \times \SO(2\ell)\, w/ \,2h \text{  vec.} ,\nonumber \\
\end{eqnarray}
labeled by two integers $\ell=1, \dots, N$ and $h=1, \dots, \lfloor \frac{F}{2} \rfloor$, parametrizing the configuration of holonomies and background for the real masses as

\begin{eqnarray}
    \label{eq:hol_SO}
    &&\begin{cases}
       \Delta_{2a} = \frac{1}{2} + \mu_a r_1 &\quad a= 1,\dots, h \\
       \Delta_{2a-1} = - \frac{1}{2} + \mu_a r_1 &\quad a=1,\dots, h \\
       \Delta_a = \mu_a r_1 &\quad a=1,\dots, F - 2h
    \end{cases}
    \nonumber \\
    &&\begin{cases}
        u_i = \frac{1}{2} + \sigma_i r_1 &\quad \quad \quad \; \; \, i=1,\dots, \ell \\
        u_i = \sigma_i r_1 &\quad \quad \quad \; \; \, i=1,\dots, F - 2h.
    \end{cases}
\end{eqnarray}

Notice that only for even $F = 2h$ the effective 3d theory at the origin of the circle can be completely lifted. As we will discuss in Section \ref{sec:APPIII}, this case corresponds to a different duality, where the monopole superpotential proportional to $Y^2$ is turned on.
Reducing the electric index on the configuration described in (\ref{eq:hol_SO}), we get

\begin{equation}
    \mathcal{I}_{\varepsilon}^{el} \underset{r_1 \to 0}{\sim}e^{\Phi_{1,2}^{el\;'}} \mathcal{Z}_{\SO(2N + \varepsilon - 2\ell)}^{2f + \varepsilon_f - 2h}
    \mathcal{Z}_{\SO(2\ell)}^{2h},
\end{equation}
with the interactions between the two partition functions constrained by the balancing conditions
\begin{equation}
        \label{BCOmu}
        \sum_{a=1}^{F} \mu_a = \omega(F-2N-\varepsilon+2).
\end{equation}
Similarly, reducing the magnetic index on the background defined in (\ref{eq:hol_SO}) with $h$ flavors and $K$ holonomies at $1/2$, we get 
\begin{equation}
    \mathcal{I}_{\varepsilon_m}^{mag} \underset{r_1 \to 0}{\sim}e^{\Phi_{1,2}^{mag\;'}} 
    \hat{\mathcal{Z}}_{\SO(2f - 2N + 4 - \varepsilon_m - 2 kl)}^{2f + \varepsilon_f - 2h}
    \hat{\mathcal{Z}}_{\SO(2K)}^{2h}.
\end{equation}
Matching the dual vacua requires tuning $K$ so that $\Phi_{1,2}^{mag\;'} - \Phi_{1,2}^{el\;'} = 0$. The difference can be written in a unique way for the four cases listed in Table \ref{tab:so}:
\begin{equation}
   \frac{i\pi(h - K - \ell - 1)\left((2h + 2K - 2\ell + \varepsilon_m ) 2\omega - 4\sum_{a =F - 2h}^{F} \mu_a \right)}{4 r_1\omega_1\omega_2} \overset{!}{=} 0.
\end{equation}

The effective 3d duality  on the partition function corresponds to the identity: 
\begin{eqnarray}
\label{O(N)onS1}
\mathcal{Z}_{\mathrm{SO}(2N+\epsilon-2\ell)}(\mu^{(1)}) \!\!\!\!&&\!\! \!\!\mathcal{Z}_{\mathrm{SO}(2\ell)}(\mu^{(2)})
=
\!\!\!\! \prod_{1\leq a \leq b\leq F - 2h} \!\!\!\! \Gamma_h(\mu_1^{(a)} +\mu_1^{(b)})  \prod_{1\leq a\leq b\leq 2h} \Gamma_h(\mu_2^{(a)}+\mu_2^{(b)}) 
\nonumber \\
&\times&
\mathcal{Z}_{\mathrm{SO}(F - 2h - 2N+2\ell -\varepsilon+ 2)}(\omega-\mu^{(1)}) \mathcal{Z}_{\mathrm{SO}(2h-2\ell+2)}(\omega-\mu^{(2)}).
\nonumber \\
\end{eqnarray}
with the balancing condition (\ref{BCOmu}).
Again we have inserted back the explicit dependence of the partition functions from the flavor fugacities and we have split
 the fugacity vectors $\mu$  as $(\mu^{(1)},\mu^{(2)})$, corresponding to the split into $F- 2h$ and $2h$ real masses
for  the vectors.

The 3d limit is obtained by first applying locally the integral identity corresponding to the pure 3d duality for orthogonal gauge theories.
It can be read from \cite{Benini:2011mf} by re-adapting it to our setup, and it is 
\begin{eqnarray}
\label{Aha3dOfromdua}
&&
 \mathcal{Z}_{\mathrm{SO}(2h-2\ell+2)}(\omega - \mu^{(2)})
 =
 \Gamma_h \left( 2 \omega (\ell-h) +\sum_{a=1}^{2h}
\mu_a^{(2)} \right)
 \nonumber \\
&&
 \prod_{1\leq a\leq b\leq 2h} \Gamma_h(2\omega-\mu_a^{(2)}-\mu_b^{(2)}) 
 \mathcal{Z}_{SO(2\ell)}(\mu^{(2)}).
\end{eqnarray}
Plugging this identity on the RHS of (\ref{O(N)onS1}) and using (\ref{BCOmu}) in the argument of the two singlets in the first line of 
(\ref{Aha3dOfromdua})
we obtain
\begin{eqnarray}
\label{O(N)3d}
&&
\mathcal{Z}_{\mathrm{SO}(2N-2\ell+\epsilon)}(\mu^{(1)}) 
=
 \Gamma_h \left(\omega(F - 2h - 2N+2\ell -\varepsilon+ 2)- \sum_{a=1}^{F-2h}
 \mu_a^{(1)} \right)
\nonumber \\
&&\prod_{1\leq a \leq b\leq F-2h} \Gamma_h(\mu_a^{(1)} +\mu_a^{(1)}) 
\mathcal{Z}_{\mathrm{SO}(F - 2h - 2N+2\ell -\varepsilon+ 2)}(\omega - \mu^{(1)}). 
\nonumber \\
\end{eqnarray}
The final step has been obtained after getting rid of the common integrals $\mathcal{Z}_{\mathrm{SO}(2\ell)}(\mu^{(2)})$.
 This indeed corresponds to consider the pure 3d limit where the physics at large distance in the Coulomb branch is effectively decoupled.
 The monopoles of the electric theory acting as singlets in the dual model are not lifted by the limit, and they appear in the final identity,
 which holds without any balancing condition. This final identity is the partition function realization of 3d  duality for orthogonal SQCD, matching with the results of \cite{Benini:2011mf}, obtained here
 from the double scaling limit of the identity between the supersymmetric indices of $\mathrm{SO}(N)$ 4d duality.

%
%
%
%
%
%
\section{3d $\SO(2N)$ SQCD and $W=Y^2$}
\label{sec:APPIII}
%
%
%
%
%
%

We conclude our analysis by reproducing a duality that was proposed in \cite{Amariti:2018gdc} from the application of the duplication formula
to the $\USp(2N)$ case in presence of monopole superpotential.
The final duality relates  $\SO(N)$ and $\SO(F-N+2)$ SQCD with $F$ vectors and a monopole superpotential $W \propto Y^2$ in both the electric and magnetic case.
There are various possible variants of such duality involving the gauging of $\mathbb{Z}_2^{C,M}$ and in some case we need to consider $Y_{\Spin}$ instead of $Y^2$.
We will omit such discussion here and concentrate to the $\SO(N)$ case.

The duality can be derived by exploiting our procedure if we consider $N=2n$ and $F=2f$ as follows. We start from 4d and fix a background for the flavor symmetry where all the flavor fugacity are set at $1/2r_1$. The gauge holonomies of both the dual phases are at $1/2r_1$ as well, we can verify that this is a vacuum only if there is an even number of vectors.

We have two possibilities: either the electric and the magnetic orthogonal gauge groups are both even, or they are both odd in 4d. In the first case the flavor background at $1/2r_1$ 
gives the same result as considering it at the origin. There is a KK monopole superpotential $W_e \propto \eta Z$ and $W_m \propto \tilde \eta \tilde Z$ in both the electric and the magnetic phase, the $Y$ direction of the Coulomb branch is unlifted, and we do not obtain any non-trivial  identity by reducing the index to the partition function in the two phases.

The situation is more interesting in the case of an odd number of colors. In such a case we have a 4d duality between $\SO(2n+1)$ and $\SO(2f-2n+3)$ SQCD both with $2f$ vectors in 4d while we arrive at a duality between $\SO(2n)$ and $\SO(2f-2n+2)$ SQCD both with $2f$ vectors on $S^1$. The interesting aspect of this duality regards the monopole superpotential that corresponds to $W_e \propto Y^2$  and $W_m \propto \tilde Y^2$.

The relation between the 4d indices in this case gives a well-defined identity for this duality that is
\begin{eqnarray}
\label{O(N)YSpin}
\mathcal{Z}_{\SO(2n)}(\mu) 
=
\prod_{1\leq a \leq b\leq 2f} \Gamma_h(\mu^{(a)} +\mu^{(b)}) 
\mathcal{Z}_{\SO(2f-2n+2)}(\omega-\mu), 
\nonumber \\
\end{eqnarray}
with the balancing condition 
\begin{equation}
\sum_{a=1}^{2f} \mu_a = 2\omega (f-n+1),
\end{equation}
where this  constraint, following from the cancellation of the axial anomaly in 4d, is imposed in 3d  by the $Y^2$ monopole in the effective superpotential when considering a finite radius $r_1$ for $S^1$.
The identity (\ref{O(N)YSpin}) corresponds to the one proven in \cite{Amariti:2018gdc} with the help of the duplication formula and here we have shown that for this specific choice of gauge ranks this effective duality can be proven from the circle compactification in presence of non-trivial flavor holonomies.

%
%
%
%
%
%
\section{Conclusions}
\label{sec:Conc}
%
%
%
%
%
%
In this paper we studied the 4d/3d reduction of $\mathcal{N}=1$ supersymmetric gauge theories by modifying the prescription of \cite{Aharony:2013dha} such to allow a double scaling limit 
by considering a small limit of the radius of the $S^1$ together with some large real mass.
The prescription given here restricts the scaling of such masses as the inverse radius, with a proportionality factor setting them to a special point of the circle, \textit{i.e.} the point mirroring the origin of the circle.
Such point is somehow special, as already observed in \cite{Aharony:2013dha} (see also \cite{Amariti:2015mva} for a geometric interpretation in terms of the Hanany-Witten setup) because it is often associated to  
a gauge symmetry enhancement. Here we have observed that many simplifications occur at such point both from the field theory perspective and from the reduction of the 4d supersymmetric index to the 3d $S^3$ partition function.
This last aspects was actually the main motivation behind our analysis. Indeed, as discussed in \cite{Aharony:2013kma} the  prescription of \cite{Aharony:2013dha} applied to the  SQCD duality with orthogonal gauge groups cannot be applied to the localization setup, because the effective description on the circle did not give rise to a finite result for the $S^3$ partition functions on both sides of the duality. 
It was then impossible to recover the expected integral identity for the orthogonal version of Aharony duality, originally worked out in \cite{Benini:2011mf}.
The double scaling considered here circumvents the problem because it removes the divergences in the effective description. This is possible because the presence of some large real mass in the small radius limit requires also a different scaling for some of the gauge holonomies. This removes some flat directions in the Coulomb branch and, 
once the prescription is translated in the language of localization, it has the net effect of removing the divergencies in the $S^3$ partition function.
In order to corroborate the result we studied also the case of unitary and symplectic SQCD, where our results are consistent with the literature.
We concluded our analysis with the case of orthogonal gauge theories where all the gauge holonomies are forced to be at the mirror point. In this case we recover a duality already proposed in the literature, either with a quadratic fundamental monopole superpotential for a $\SO(2n)$gauge group or with a $Y_{\Spin}$ monopole superpotential 
 for a $\Spin(2n)$ gauge group.

There are many open questions and possible lines of investigations that we left to future analysis.

First, an interesting follow-up of our analysis consists of studying type $A$ and $D$ superpotentials for orthogonal gauge theories, reducing the dualities of \cite{Leigh:1995qp,Intriligator:1995ff,Intriligator:1995ax} and \cite{Brodie:1996vx,Brodie:1996xm}
respectively. Such cases can give origin to more intricate structure of monopole superpotential and to web of dualities.
While we are planning to be back to this problem in the future we have already verified that our procedure applied to such cases is consistent with the results 
of \cite{Amariti:2015mva} obtained from the brane setup, in the case of type $A$ superpotential.

Another interesting aspect that we did not investigate is associated to the analysis of section \ref{sec:APPIII}. Indeed, the duality with the quadratic monopole superpotential proposed there was already discussed in \cite{Amariti:2019rhc} following from the brane setup. Such case regarded an affine compactification of a 4d duality involving SQCD with $so(2n+1)$ algebra.
While the affine compactification of $so(2n)$ and of $usp(2n)$ do not give rise to any new duality when considering the holonomies at the mirror point on the circle, this case
is special because a new duality emerges. This is due, from the brane picture perspective, to the fate of the $O4$ plane under T-duality. 
On the other hand other possible compactifications are possible, by allowing a further twist by an outer automorphism of the gauge algebra. Such twisted affine compactifications have been investigated from the brane perspective in \cite{Amariti:2019rhc} and it should be interesting to study them following the prescription proposed here.

As we stressed above we did not study other possible large real mass scalings with respect to the small radius of the circle. 
In general this possibility is expected to give origin to 3d dualities with special unitary or unitary gauge groups and non-trivial monopole superpotentials of the type proposed in \cite{Benini:2017dud,Amariti:2018gdc,Amariti:2019rhc}. It would be interesting to 
derive these results along the lines of the 4d/3d reduction in presence of a double scaling on the real masses.

\section*{Acknowledgments}
We are grateful to Simone Rota for discussions. The work of the authors has been supported in part by the Italian Ministero dell’Istruzione, Università e Ricerca (MIUR), in part by Istituto Nazionale di Fisica Nucleare (INFN) through the “Gauge Theories, Strings, Supergravity” (GSS) research project.

\bibliographystyle{JHEP}
\bibliography{biblio.bib}
\end{document}